\documentclass[fleqn,usenatbib,usedcolumn]{mnras}

\pdfoutput=1

\usepackage{txfonts}

\usepackage[T1]{fontenc}
\usepackage{ae,aecompl}

\usepackage{graphicx}	
\usepackage[usenames,dvipsnames]{color}

\newcommand{\Sref}[1]{Section \ref{#1}}
\newcommand{\Tref}[1]{Table \ref{#1}}

\sfcode`\.=1001\sfcode`\?=1001\sfcode`\!=1001
\newcommand{\Fref}[1]{\ifhmode \ifnum\spacefactor=1001 Figure \ref{#1}\else Fig.\ \ref{#1}\fi \else Figure \ref{#1}\fi}
\newcommand{\Eref}[1]{\ifhmode \ifnum\spacefactor=1001 Equation (\ref{#1})\else equation (\ref{#1})\fi \else Equation (\ref{#1})\fi}

\newcommand{\ms}{\ensuremath{\textrm{m\,s}^{-1}}}
\newcommand{\kms}{\ensuremath{\textrm{km\,s}^{-1}}}

\newcommand{\pcmsq}{\ensuremath{\textrm{cm}^{-2}}}
\newcommand{\SN}{\ensuremath{\textrm{S/N}}}

\newcommand{\zem}{\ensuremath{z_\textrm{\scriptsize em}}}
\newcommand{\zab}{\ensuremath{z_\textrm{\scriptsize abs}}}

\newcommand{\Nion}[2]{\ensuremath{N_{\rm #1\textsc{\scriptsize{\,#2}}}}}
\newcommand{\lNHI}{\ensuremath{\log(N_\textsc{h\scriptsize{\,i}}/\textrm{cm}^{-2})}}
\newcommand{\Ion}[2]{\ensuremath{\textrm{#1\,{\scshape{#2}}}}}
\newcommand{\tran}[3]{\ensuremath{\Ion{#1}{#2}\,\lambda\textrm{#3}}}

\newcommand{\ewr}[3]{\ensuremath{W_\textrm{\scriptsize r}(\Ion{#1}{#2}\,\lambda{#3})}}
\newcommand{\daa}{\ensuremath{\Delta\alpha/\alpha}}

\usepackage{array}
\newcolumntype{:}{>{\global\let\currentrowstyle\relax}}
\newcolumntype{;}{>{\currentrowstyle}}
\newcommand{\rowstyle}[1]{\gdef\currentrowstyle{#1}%
  #1\ignorespaces
}

\title[Varying $\alpha$ with Zn and Cr]{Precise limits on cosmological variability of the fine-structure constant with zinc and chromium quasar absorption lines}

\author[M. T. Murphy, A. L. Malec, J. X. Prochaska]{Michael
  T. Murphy,$^{1}$\thanks{E-mail: mmurphy@swin.edu.au (MTM)} Adrian
  L. Malec,$^{1}$ J. Xavier Prochaska$^{2}$\\
  $^{1}$Centre for Astrophysics and Supercomputing, Swinburne University of Technology, Hawthorn, Victoria 3122, Australia\\
  $^{2}$Department of Astronomy and Astrophysics, UCO/Lick Observatory, University of California, 1156 High Street, Santa Cruz, CA 95064, USA
}

\date{Accepted 2016 June 16. Received 2016 June 16; in original form 2016 May 26}

\pubyear{2016}

\volume{{\rm 461}}

\begin{document}
\label{firstpage}
\pagerange{\pageref{firstpage}--\pageref{lastpage}}
\maketitle

\begin{abstract}
The strongest transitions of Zn and \Ion{Cr}{ii} are the most sensitive to relative variations in the fine-structure constant (\daa) among the transitions commonly observed in quasar absorption spectra. They also lie within just 40\,\AA\ of each other (rest frame), so they are resistant to the main systematic error affecting most previous measurements of \daa: long-range distortions of the wavelength calibration. While Zn and \Ion{Cr}{ii} absorption is normally very weak in quasar spectra, we obtained high signal-to-noise, high-resolution echelle spectra from the Keck and Very Large Telescopes of 9 rare systems where it is strong enough to constrain \daa\ from these species alone. These provide 12 independent measurements (3 quasars were observed with both telescopes) at redshifts 1.0--2.4, 11 of which pass stringent reliability criteria. These 11 are all consistent with $\daa=0$ within their individual uncertainties of 3.5--13\,parts per million (ppm), with a weighted mean $\daa = 1.2\pm1.7_{\rm stat}\pm0.9_{\rm sys}$\,ppm (1$\sigma$ statistical and systematic uncertainties), indicating no significant cosmological variations in $\alpha$. This is the first statistical sample of absorbers that is resistant to long-range calibration distortions (at the $<$1\,ppm level), with a precision comparable to previous large samples of $\sim$150 (distortion-affected) absorbers. Our systematic error budget is instead dominated by much shorter-range distortions repeated across echelle orders of individual spectra.
\end{abstract}




\section{Introduction}\label{s:intro}

Quasar absorption spectra from 8--10-m optical telescopes have
provided the most precise constraints on variations in the
fine-structure constant, $\alpha\equiv e^2/\hbar c$, on cosmological
scales. The high resolving power ($R\ga40000$) and high
signal-to-noise ratio ($\SN\ga50$\,per resolution element) provided by
echelle spectrographs on these telescopes allows the (usually) complex
velocity structures of metal absorption lines to be resolved and
compared between transitions via detailed profile fitting
techniques. The introduction of the many-multiple (MM) method by
\citet{Webb:1999:884} and \citet{Dzuba:1999:888} enabled a further
increase in precision: the detailed pattern of velocity shifts between
many different metal transitions, caused by a possible variation in
$\alpha$, is fully utilised, providing additional sensitivity and
increased statistics compared to previous analyses of alkali doublets
\citep[such as from \Ion{C}{iv}, \Ion{Mg}{ii}, \Ion{Si}{iv} etc.;
e.g.][]{Bahcall:1967:L11,Wolfe:1976:179,Cowie:1995:596,Varshalovich:1996:6,Murphy:2001:1237}. For
each transition, $i$, analysed in a quasar absorption spectrum, its
response to a relative $\alpha$-variation, $\Delta\alpha/\alpha$, is
characterised by its sensitivity coefficient, $q_i$, resulting in a
velocity shift, $\Delta v_i$:
\begin{equation}\label{e:daa}
\daa\equiv\frac{\alpha_z-\alpha_0}{\alpha_0}\approx-\frac{1}{2}\frac{\Delta v_i}{c}\frac{\omega_i}{q_i}\equiv\frac{1}{2}\frac{\Delta v_i}{c}\frac{1}{Q_i}\,,
\end{equation}
where $\omega_i$ is transition $i$'s wavenumber. Therefore,
together with the \SN\ of the spectrum and the sharpness of the
absorption features, the spread in $Q$ coefficients among the
transitions analysed defines the precision with which
$\Delta\alpha/\alpha$ can be measured.

Early MM studies focussed on very large quasar absorber samples from
archival spectra -- those not observed or calibrated specifically for
measuring \daa\ -- and maximised precision by comparing as many strong
transitions as possible in each absorber. Interestingly, spectra from
the Keck telescope's High Resolution Echelle Spectrometer (HIRES)
indicated some evidence for a smaller $\alpha$ at redshifts
$\zab=0.2$--4.2 than the current laboratory value
\citep{Webb:1999:884,Murphy:2001:1208}, with the final sample of 143
absorbers indicating $\daa=-5.7\pm1.1$\,parts per million
\citep[ppm;][]{Murphy:2003:609,Murphy:2004:131}. Spectra from the Very
Large Telescope's (VLT's) Ultraviolet and Visual Echelle Spectrograph
(UVES) did not replicate this result \citep[][though see
\citealt{Murphy:2008:1053,Wilczynska:2015:3082}]{Chand:2004:853}, with
the most recent large sample of 153 absorbers indicating
$\daa=+2.1\pm1.2$\,ppm \citep{King:2012:3370}. Nevertheless, when
combined, the large HIRES and UVES samples supported self-consistent,
4$\sigma$ evidence for a coherent, dipole-like variation in $\alpha$
across the sky \citep[][]{Webb:2011:191101,King:2012:3370}. Studies of
specific astrophysical or instrumental systematic effects could not
explain these surprising results
\citep[e.g.][]{Murphy:2001:1223,Murphy:2003:609,King:2012:3370}. Several
detailed studies of very high \SN\ spectra of individual quasars have
also reported \daa\ measurements with $\ga$3\,ppm uncertainties
\citep[e.g.][]{Quast:2004:L7,Levshakov:2005:827,Levshakov:2007:1077,Molaro:2008:173,Murphy:2008:1053,Agafonova:2011:28,Molaro:2013:A68,Songaila:2014:103}
which were consistent with no $\alpha$-variation but which did not
rule out the large-sample results.

The most important systematic effects in this context are those that
can distort the wavelength scale, inducing spurious velocity shifts
\emph{between} transitions at different wavelengths [cf.\ the same
shift in all transitions; see \Eref{e:daa}]. Given the surprising
results above, such distortions from instrumental effects are of
paramount concern. Empirical tests that are sensitive to all such
effects, whether from known or, most importantly, unknown causes are
therefore extremely valuable. Particularly successful has been the
comparison of the relative wavelength scales established using the
standard thorium--argon (ThAr) lamp technique, which was used in the
quasar absorption studies, with a more accurate one embedded in the
spectrum of the object itself. \citet{Molaro:2008:559} first compared
the centroid wavelengths of discrete lines in solar atlases with
those measured in ThAr-calibrated asteroid spectra from UVES, finding
no long-range distortions of the ThAr wavelength scale. Similarly,
\citet{Griest:2010:158} and \citet{Whitmore:2010:89} observed quasars
through an iodine gas cell with HIRES and UVES, respectively, and
found no long-range distortions (though they did identify short-range
distortions -- see \Sref{sss:intra}). These early
``super-calibration'' studies therefore provided some confidence that
the ThAr-calibrated wavelength scales in the previous quasar
absorption studies were accurate enough for reliable constraints on
\daa.

Unfortunately, the spectra used in these early super-calibration
studies appear to have been exceptions to the
rule. \citet{Rahmani:2013:861} cross-correlated several UVES asteroid
spectra, taken between 2006 and 2012, with more accurate solar spectra
recorded with Fourier-transform spectrometers (FTSs), and found
long-range distortions up to 0.7\,\ms\,\AA$^{-1}$, enough to cause
$\sim$10\,ppm systematic effects in \daa. \citet{Bagdonaite:2014:10}
found smaller, though still substantial, distortions with a similar
analysis of asteroid and `solar-twin' stellar
spectra. \citet{Songaila:2014:103} also identified evidence for
distortions by using telluric features in several quasar
spectra. Recently, \citet{Whitmore:2015:446} super-calibrated HIRES
and UVES with archival asteroid and solar-twin spectra observed over
two decades, finding that 0.2\,\ms\,\AA$^{-1}$ distortions are
ubiquitous in these spectrographs. Furthermore, they found that these
likely explain the non-zero \daa\ results from the large UVES sample
of \citet{Webb:2011:191101} and \citet{King:2012:3370}, and at least
partially explain the earlier HIRES results of \citet{Webb:1999:884}
and \citet{Murphy:2004:131}. Therefore, it is likely that all \daa\
results derived from UVES or HIRES spectra before 2014 are
substantially affected by systematic errors due to long-range
distortions. Most of these spectra cannot be corrected retrospectively for lack of
solar-like spectra observed on the same nights as the quasar
exposures.

\citet{Evans:2014:128} measured \daa\ in 3 absorbers
towards a single bright quasar using 3 different spectrographs: HIRES,
UVES and the Subaru telescope's High Dispersion Spectrograph (HDS). By
comparing these spectra with each other, and super-calibrating them
with contemporaneous asteroid and stellar iodine-cell observations,
they corrected them for long-range distortions and obtained a weighted
mean \daa\ of $-5.4\pm3.3_{\rm stat}\pm1.5_{\rm sys}$\,ppm at
$\zab=1.1$--1.8. This is the only distortion-corrected measurement of
\daa\ published so far \citep[though see \Sref{s:disc} for discussion
of a new, even more precise measurement in a single absorber
by][]{Kotus:2016:arXiv}. It is therefore vital to supplement this with
a substantial number of new measurements which either avoid or remove
the long-range distortions evident in all three slit-based
spectrographs on 8--10-m telescopes mentioned above.

\begin{figure}
\begin{center}
\includegraphics[width=1.0\columnwidth]{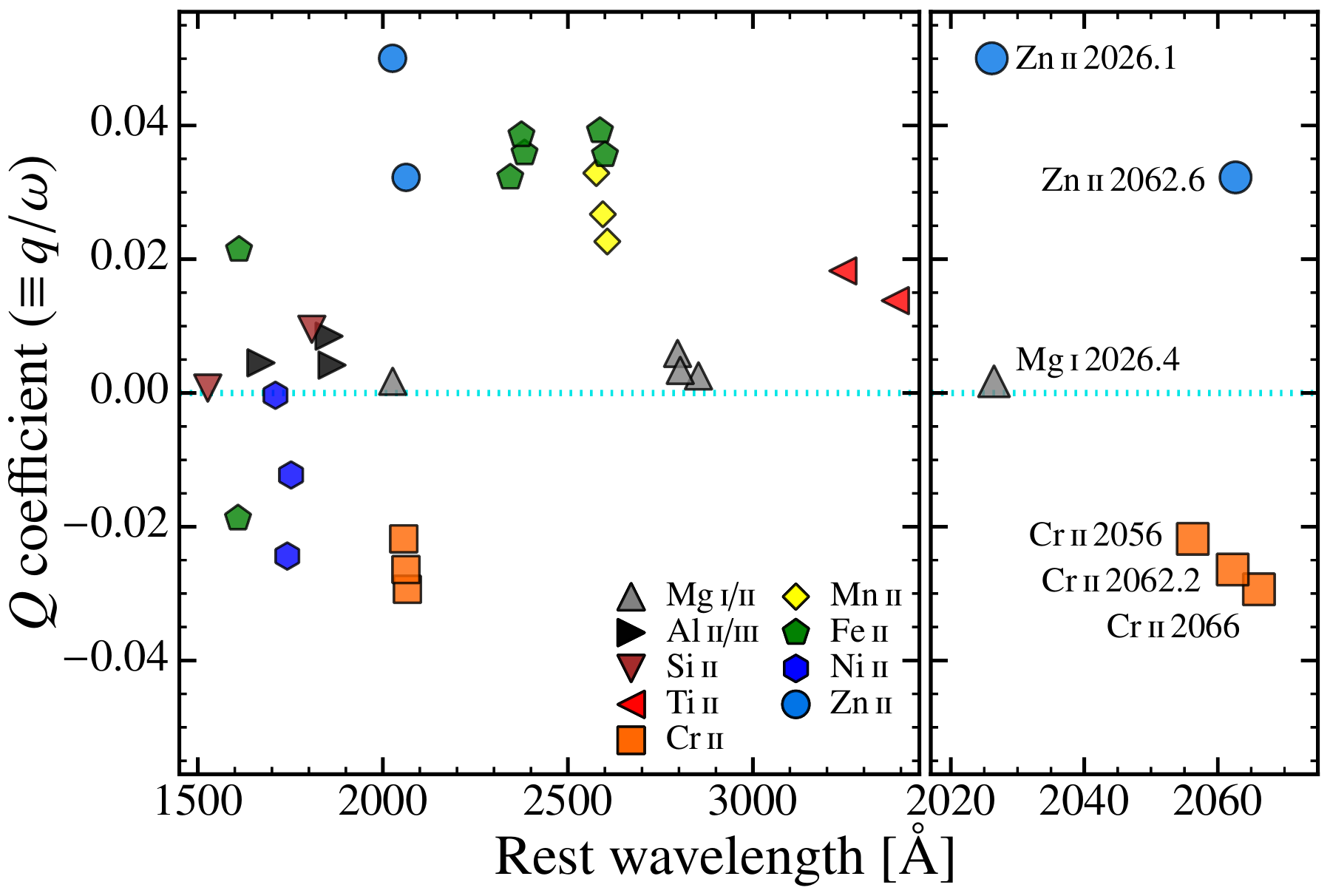}\vspace{-1em}
\caption{Left panel: The sensitivity of transition frequencies to
  variation in $\alpha$, represented by the $Q$ coefficient, defined as $Q\equiv q/\omega$ [see
  \Eref{e:daa}], for transitions used in MM analyses to date. Right
  panel: The small wavelength region in which the \Ion{Zn/Cr}{ii}
  transitions that are the focus of this paper fall, showing their
  large $Q$ values with opposing sign. The $q$ values, laboratory
  wavelengths and other atomic data reviewed in
  \citet{Murphy:2014:388} are used in throughout this work. The
  uncertainties in $Q$ are $\sim$0.5 and $1.4\times10^{-3}$ for the Zn
  and \Ion{Cr}{ii} transitions, respectively, which are negligible for
  this work.}
\label{f:q_vs_wl}
\end{center}
\end{figure}

In this paper we seek to avoid the effects of long-range distortions
by using only the closely-spaced \Ion{Zn}{ii} and \Ion{Cr}{ii}
transitions in an MM analysis. The principle behind this approach is
evident in \Fref{f:q_vs_wl} which shows the $Q$ coefficients of the
transitions used in MM analyses to date, as reviewed in
\citet{Murphy:2014:388}. The \Ion{Zn}{ii}\,$\lambda\lambda$2026/2062
doublet transitions have some of the largest \emph{positive} $Q$
coefficients, while the very nearby
($\Delta\lambda_{\rm rest}\le$40\,\AA) \Ion{Cr}{ii} triplet
transitions ($\lambda$2056, 2062, 2066) have amongst the largest
\emph{negative} $Q$ values. That is, the Zn and \Ion{Cr}{ii}
transitions shift by large amounts, but in opposite directions if
$\alpha$ varies, despite their very small wavelength
separation. Therefore, analysing only the \Ion{Zn/Cr}{ii} combination
simultaneously provides high sensitivity to variations in $\alpha$ and
low sensitivity to long-range distortions of the wavelength
scale. Indeed, for an absorber at redshift $\zab=1$, a typical
0.2\,\ms\,\AA$^{-1}$ distortion would only cause a spurious
$\approx$16\,\ms\ shift between the \tran{Zn}{ii}{2026} line and the
\Ion{Cr}{ii} triplet, corresponding to a systematic error in \daa\ of
just $\approx$0.3\,ppm [via \Eref{e:daa}], substantially smaller than the
statistical errors from the large HIRES and UVES samples or
individual, high \SN\ absorbers. Note also that the
\tran{Zn}{ii}{2062} transition falls in between the
\tran{Cr}{ii}{2062} and $\lambda$2066 transitions, producing a more
detailed, more characteristic pattern of velocity shifts due to a
varying $\alpha$. This increases resistance to long-range distortions
as well as short-range distortions (see further discussion in Sections
\ref{ss:syserr} \& \ref{s:disc}).

Here we obtain 12 independent MM measurements of \daa\ with this
\Ion{Zn/Cr}{ii} approach, using both new and archival quasar spectra
from HIRES and UVES. The observations and data reduction steps are
described in \Sref{s:obs}, while \Sref{s:anal} presents our approach
to fitting the absorption profiles to measure \daa, and our analysis
of each absorption system. The \daa\ results and systematic errors are
analysed in \Sref{s:res}. The results are compared to others in the
recent literature in \Sref{s:disc}, where we also compare the main
advantages and disadvantages of the \Ion{Zn/Cr}{ii}
approach. \Sref{s:conc} summarizes our conclusions.

\section{Quasar observations and data reduction}\label{s:obs}

\subsection{Quasar selection}\label{ss:qsosel}

Our main aim is to obtain a precision in \daa\ that is competitive
with previous constraints, $\sim$2\,ppm, from an ensemble of MM
measurements using just the \Ion{Zn/Cr}{ii} transitions. The
statistical precision in \daa\ available from an individual absorption
system varies considerably according to several factors, particularly
the \SN\ of the spectrum, and the optical depths and velocity
structure of the metal absorption lines. Deeper absorption (though
preferably unsaturated), with more distinct, narrow spectral features,
provides more spectral information with which velocity
shifts between transitions can be determined. However, the Zn and \Ion{Cr}{ii}
absorption in most absorption systems is very weak, presenting
(rest-frame) \Ion{Zn}{ii} equivalent widths of
$\ewr{Zn}{ii}{2026}\la0.3$\,\AA. Even amongst the relatively high
column-density damped Lyman-$\alpha$ systems [i.e.\ those with
$\lNHI>20.3$\,\pcmsq], ``metal-strong'' systems that exceed this
equivalent-width limit account for only $\sim$5\,per cent of the
population \citep{Herbert-Fort:2006:1077}. And, of course, even
amongst these ``metal-strong'' systems, few occur along lines of sight
towards quasars bright enough to allow high-\SN\ echelle-resolution
spectra to be easily obtained.

Therefore, to establish a sample of MM measurements with maximal
statistical precision in \daa\ and minimal additional telescope
observing time, we selected targets for which existing UVES and/or
HIRES spectra showed strong Zn and \Ion{Cr}{ii} absorption lines,
preferably with complex velocity structure with many distinct, narrow
features. Weaker, less complex absorbers were included only when the
quasar was significantly brighter, allowing a higher spectral \SN\ to
be obtained so that a reasonably uniform precision on \daa\ per
absorber could be expected. We identified the 9 quasars that are
listed in \Tref{t:obs}, each with one absorber along its line of
sight, that satisfied these criteria. The quasars have magnitudes
ranging from $r=16.1$ to 18.3\,mag, emission redshifts
$\zem=1.4$--2.7 and the absorbers fall in the redshift range
$\zab=1.0$--2.4.

\begin{table*}
\begin{center}
  \caption{Observational details of the quasar sample. The first
    column is the quasar name, where H or U superscripts denote
    Keck/HIRES or VLT/UVES observations, respectively, with the J2000
    right-ascension and declination, and the quasar emission (\zem)
    and absorption (\zab) redshifts provided. The spectrograph
    wavelength setting is specified as the approximate wavelength
    range for HIRES and the setting name (according to approximate
    central wavelength) for UVES. Total exposure times are specified
    for groups of exposures (specified in parentheses) in the same
    setting on each row; HIRES exposures of a given quasar were
    usually taken in the same setting, whereas a range of settings was
    often employed with UVES. The slit-widths correspond to the
    nominal resolving powers specified in the notes$^{a}$. The final
    two columns give the total \SN\ per pixel, after all exposures are
    combined for each quasar, at rest-wavelengths
    ($\lambda_{\rm rest}$) near the \Ion{Cr}{ii} and
    \tran{Mg}{i}{2852} transitions. The on-chip binning for the UVES
    observations was 2$\times$2, except for those of J0058$+$0041
    where it was 1$\times$1 (i.e.\ unbinned). All HIRES observations
    were binned only in the spatial direction (i.e.\ 2$\times$1
    binning). The dispersions of the combined spectra were set to 1.3
    and 2.5\,\kms\ for unbinned and binned spectra,
    respectively. Program identifiers, principle investigator surnames,
    observing mode (visitor or service) and dates are given in the
    notes for each observing run. The red cross-disperser was used for
    all HIRES observations except those taken in runs I, J \& K.}
\label{t:obs}
{\footnotesize
\begin{tabular}{lccccclccc}\hline
Quasar & RA [hr] & Dec.\ [deg] & \zem\ & \zab & Setting & Exposure time & Slit-width$^{\rm a}$ & \multicolumn{2}{c}{\SN\,pix$^{-1}$ at $\lambda_{\rm rest}\sim$} \\
       & \multicolumn{2}{c}{(J2000)} & &      & [nm]    & [s]           & [arcsec]            & 2060\,\AA & 2852\,\AA \\\hline
J0058$+$0041$^{\rm H}$ & 00:58:24.75 & $+$00:41:13.6 & 1.92 & 1.072 & 379--829$^{\rm b}$  & 37200 (11\,$\times$\,3300--3600)$^{\rm C,E,F}$ & 0.86                & 50 &  68 \\
J0058$+$0041$^{\rm U}$ &             &               &      &       & 390$+$564 & 17300 (6\,$\times$\,2886)$^{\rm L}$              & 0.8                 & 29 &  31 \\ 
                       &             &               &      &       & 470$+$760 & 12900 (4\,$\times$\,3000--3600)$^{\rm S}$        & 1.0                 &    &     \\ 
PHL957$^{\rm H}$       & 01:03:11.27 & $+$13:16:17.7 & 2.68 & 2.309 & 424--869  & 31620 (11\,$\times$\,1200--3600)$^{\rm A}$       & 0.86$^{\rm c}$      & 74 & --- \\
PHL957$^{\rm U}$       &             &               &      &       & 390$+$860 &  7200 (2\,$\times$\,3600)$^{\rm O}$              & 1.0$+$0.9$^{\rm d}$ & 95 & --- \\ 
                       &             &               &      &       & 390$+$590 & 15400 (3\,$\times$\,5000--5400)$^{\rm Q}$        & 0.9$^{\rm e}$       &    &     \\ 
J0108$-$0037$^{\rm U}$ & 01:08:26.84 & $-$00:37:24.2 & 1.37 & 1.371 & 390$+$580 & 14800 (4\,$\times$\,3700)$^{\rm T}$              & 1.0                 & 48 &  58 \\ 
                       &             &               &      &       & 346$+$580 &   860 (2\,$\times$\,430)$^{\rm V}$               & 1.0                 &    &     \\  
                       &             &               &      &       & 437$+$860 &   860 (2\,$\times$\,430)$^{\rm V}$               & 1.0                 &    &     \\  
J0226$-$2857$^{\rm U}$ & 02:26:20.50 & $-$28:57:50.8 & 2.17 & 1.023 & 390$+$564 & 12800 (4\,$\times$\,2900--3300)$^{\rm M}$        & 0.8                 & 35 &  46 \\ 
                       &             &               &      &       & 390$+$580 &  9000 (3\,$\times$\,3000)$^{\rm R}$              & 1.2                 &    &     \\  
                       &             &               &      &       & 437$+$760 & 13560 (4\,$\times$\,2460--4200)$^{\rm M}$        & 0.8                 &    &     \\  
J0841$+$0312$^{\rm H}$ & 08:41:06.78 & $+$03:12:06.6 & 1.94 & 1.342 & 414--848  & 26100 (9\,$\times$\,2400--3600)$^{\rm C,F}$      & 0.86                & 85 & 100 \\
J0841$+$0312$^{\rm U}$ &             &               &      &       & 470$+$760 & 11580 (4\,$\times$\,2896)$^{\rm L}$              & 0.8                 & 41 &  72 \\
J1029$+$1039$^{\rm H}$ & 10:29:04.15 & $+$10:39:01.6 & 1.79 & 1.622 & 435--869  & 21600 (6\,$\times$\,3600)$^{\rm C,D,F}$          & 0.86                & 36 &  29 \\
J1237$+$0106$^{\rm H}$ & 12:37:24.51 & $+$01:06:15.4 & 2.02 & 1.305 & 400--848$^{\rm f}$  & 23600 (8\,$\times$\,1600--3600)$^{\rm C,D,F}$  & 0.86                & 30 &  36 \\ 
Q1755$+$57$^{\rm H}$   & 17:56:03.63 & $+$57:48:48.0 & 2.11 & 1.971 & 435--869  & 31200 (9\,$\times$\,3300--3600)$^{\rm B,D}$      & 0.86                & 29 &  34 \\
                       &             &               &      &       & 318--605  & 23200 (4\,$\times$\,4400--5400)$^{\rm G,I,J,K}$  & 0.86                &    &     \\
Q2206$-$1958$^{\rm U}$ & 22:08:52.07 & $-$19:44:00.0 & 2.56 & 1.921 & 346$+$580 &  9000 (2\,$\times$\,4500)$^{\rm P}$              & 1.0                 & 97 &  25 \\
                       &             &               &      &       & 390$+$564 & 20700 (6\,$\times$\,2700--3600)$^{\rm N}$        & 1.0                 &    &     \\
                       &             &               &      &       & 437$+$860 &  4200 (1\,$\times$\,4200)$^{\rm N}$              & 1.0                 &    &     \\
                       &             &               &      &       & 455$+$850 & 19600 (4\,$\times$\,4200--5400)$^{\rm N}$        & 0.9                 &    &     \\
\hline
\end{tabular}
}
\begin{minipage}{\textwidth}
{\footnotesize
Notes: $^{\rm a}$The nominal resolving powers corresponding to the HIRES 0.86 \& 1.15\,arcsec slits are $R=50000$ \& 37000, respectively. For UVES, a 1.0\,arcsec slit corresponds to a resolving power of $R=42000$ and 39000 in the blue and red arms, respectively, with proportionately higher (lower) $R$ for narrower (wider) slits.
$^{\rm b}$One 3300-s exposure was taken in the 336--810-nm setting. 
$^{\rm c}$A 1.15\,arcsec slit was used for one 3300-s exposure due to poor seeing conditions.
$^{\rm d}$1.0\,arcsec for the blue arm, 0.9\,arcsec for the red arm.
$^{\rm e}$One 5400-s exposure was taken with a 1.0\,arcsec slit.
$^{\rm f}$One 1600-s exposure was taken in the 435--869-nm setting.\\
HIRES observing programs (all in visitor mode):
$^{\rm A}$CS280Hr (Murphy): 20--22 Aug.~2008.
$^{\rm B}$CS280Hr (Murphy): 07 Aug~2008.
$^{\rm C}$CS193Hr (Murphy): 01 Feb.~2009.
$^{\rm D}$W015Hr (Murphy): 19 May 2009.
$^{\rm E}$W009Hr (Murphy): 03 Nov.~2009.
$^{\rm F}$G914Hr (Malec): 29 Dec.~2009.
$^{\rm G}$U11H (Prochaska): 09 Sep.~2004.
$^{\rm I}$U152Hb (Prochaska): 02 Jun.~2006.
$^{\rm J}$U157Hb (Prochaska): 03 Jun.~2006.
$^{\rm K}$U080Hb (Prochaska): 18--19 Aug.~2006.\\
UVES observing programs:
$^{\rm L}$084.A-0136 (Malec, service): 12, 17 \& 18 Oct.~2009 for J0058$+$0041; 15 \& 27 Dec.~2009, 09 Jan.~2010 for J0841$+$0312.
$^{\rm M}$084.A-0136 (Malec, visitor): 8 \& 9 Dec.~2009.
$^{\rm N}$65.O-0158 (Pettini, visitor): 29--31 May 2000.
$^{\rm O}$67.A-0022 (D'Odorico, service): 16 \& 17 Sep.~2001.
$^{\rm P}$072.A-0346 (Ledoux, visitor): 29 Oct.~2003.
$^{\rm Q}$074.A-0201 (Srianand, visitor): 8 \& 9 Oct.~2004.
$^{\rm R}$079.A-0600 (Bouch\'e, service): 25 \& 28 Jul.~2007, 5 Sep.~2007.
$^{\rm S}$082.A-0682 (Dessauges-Zavadsky, visitor): 05 May 2008.
$^{\rm T}$082.A-0569 (Srianand, service): 21, 23 \& 25 Nov.~2008, 03 Dec.~2008.
$^{\rm V}$083.A-0874 (Miniati, service): 20 \& 24 Aug.~2009.
}
\end{minipage}

\end{center}
\end{table*}

\subsection{Observations}\label{ss:obs}

\Tref{t:obs} lists the observational details for the 9 quasars and
absorbers selected for this work. The spectra originate from both
Keck/HIRES \citep{Vogt:1994:362} and VLT/UVES \citep{Dekker:2000:534},
and three quasars have spectra from both telescopes/spectrographs; we
analyse the HIRES and UVES spectra separately in these
cases. Therefore, our final sample effectively comprises 9 absorbers
with 12 independent \daa\ measurements.

The observations in \Tref{t:obs} include $\approx$63\,h of new
observations specifically for this work and $\approx$38\,h of
archival exposures. Of the 54\,h of HIRES exposures in total,
48\,h (90\,per cent) are new observations, with only Q1755$+$57
having a (6\,hr) contribution from archival spectra. For UVES, 70\,per
cent (32\,h) of the 47\,h of exposures are archival, while most
UVES exposures of J0058$+$0041, J0226$-$2857 and J0841$+$0312 are
new. For all quasars, multiple exposures were observed over several
nights, split across several observing runs in most cases. Both the
faintest quasar in our sample, Q1755$+$57, and also Q2206$-$1958 with
its well-studied damped Lyman-$\alpha$ system at $\zab=1.921$, have
the most exposures (13), while our UVES observations of the brightest
target, J0841$+$0312, comprise the fewest (4). The sky transparency
and seeing conditions varied considerably throughout the many
different observing runs/nights. The seeing
full-width-at-half-maximum (FWHM) was rarely less than 0.7\,arcsec. The slit width
used for all except one HIRES exposure was 0.86\,arcsec, corresponding to
a nominal resolving power of $R\approx50000$ (obtained from the ThAr
lamp exposures). For UVES, a larger variety of slit-widths was employed, but for all but 3 exposures a
0.8--1.0\,arcsec slit was used, corresponding
to (ThAr-based) nominal resolving powers $R=42000$ down to 39000, respectively. New
observations were conducted in `grey' lunations, except for the
brightest three quasars (J0058$+$0041, PHL957 and J0841$+$0312) where
bright lunations were also used.

During new observations in visitor mode\footnote{`Visitor-mode'
  observations were conducted by the authors in person, whereas
  `service-mode' observations were conducted entirely by observatory
  staff.} careful attention was paid to keeping the quasar centred
within the slit by small adjustments to the telescope tracking. This
was especially the case when only `slit guiding' was possible in the
absence of a nearby guide-star (mostly for HIRES), but this approach
was also followed when `offset guiding' from a nearby star was
possible (mostly for UVES). A difference in quasar--slit alignment
between two exposures of the same quasar will manifest itself as a
velocity shift between corresponding absorption features in the two
exposures. Indeed, HIRES observations by \citet{Evans:2014:128} of a
very bright quasar, for which these `slit shifts' can be measured
accurately, show that they are typically $\approx$200\,\ms,
corresponding to alignment differences between exposures of
$\approx$4\,per cent of the slit width. However, their observations of
the same quasar with VLT/UVES and Subaru/HDS included small telescope
tracking adjustments and this reduced the typical slit shift down to
$\approx$100\,\ms. The slit shift will, to a good approximation, be
the same at all wavelengths, so no substantial distortion is
introduced that would lead to systematic errors in \daa\ [see
\Eref{e:daa}].

However, a complicating factor arises when exposures which were taken
in different spectrograph settings are combined. A `setting shift'
arises due to a difference between the average slit shifts for the
exposures in two different settings. In the combined spectrum, the
average slit shift then effectively varies with wavelength, i.e.\
setting shifts can create a distortion that may cause a systematic
error in \daa. However, by minimizing the slit shifts in our new
observations, we also minimize the setting shifts. \Tref{t:obs} also
shows that we restricted all our new observations (except for
J0226$-$2857) to a single wavelength setting for each quasar. However, even for the
quasars with multiple settings (mostly those with contributions from
archival spectra) the effect of setting shifts is negated by our
\Ion{Zn/Cr}{ii} approach to measuring \daa: the Zn and \Ion{Cr}{ii}
transitions lie so close together that, in all except one of our spectra,
they are all covered by the same exposures. Therefore, the effect of
slit and setting shifts in our \daa\ measurements is effectively
avoided. The exceptional case is Q1755$+$57, where the 318--605-nm
wavelength setting, used in 4 exposures, covers only the
\tran{Zn}{ii}{2026} transition; the other 9 exposures cover all the Zn
and \Ion{Cr}{ii} transitions. We discuss this exception in
\Sref{sss:Q1755} and find that it is unlikely to have caused a
significant systematic error. Note that, as discussed in
\Sref{ss:genfit}, our profile fitting analysis does incorporate
velocity structure information from the \tran{Mg}{i}{2852} transition
which is often not covered by the same exposures as the
\Ion{Zn/Cr}{ii} lines. However, the velocity of that transition is
de-coupled from the others in our fitting procedure so that setting
shifts do not significantly affect our \daa\ measurements (see
\Sref{ss:genfit} for details).

Establishing the wavelength scale of the quasar exposures is the most
important calibration step for our \daa\ measurements. Therefore, in
all our new observations, a ThAr comparison lamp exposure was
`attached' to each quasar exposure, i.e.\ taken immediately after the
quasar exposure without any changes to other spectrograph settings. We
also restricted our selection of archival exposures to those that were
either attached or, in some cases on UVES, taken immediately before
the quasar exposure in the same setting. In the latter cases, the
grating encoders will have been re-initialized between the ThAr and
quasar exposure, potentially leading to velocity shifts between
different exposures of the same quasar, though probably not
substantial distortions; that is, this is effectively an additional
slit shift that will, by the same argument as above, not strongly
influence our \daa\ measurements. Similarly, for two pairs of
exposures for each of J0108$-$0037, Q1755$+$57 and Q2206$-$1958, the
ThAr calibration exposure was attached to the later of the pair of
quasar exposures. Again, this is unlikely to affect \daa\
appreciably. Finally, for UVES archival spectra in the very red 760,
850 and 860-nm settings, the ThAr exposures were taken separately at
the end of the night. The only transition used in our analysis from
these settings is \tran{Mg}{i}{2852} but, as described above, our
fitting approach means that this will not affect the \daa\
measurements.

\subsection{Data reduction}\label{ss:reduce}

All UVES exposures were reduced with the ESO Common Pipeline Language
data-reduction
software\footnote{{\urlstyle{rm}\url{http://www.eso.org/observing/dfo/quality/UVES/pipeline/pipe_reduc.html}}}. The
general reduction scheme and the specific procedures used here were
the same as those described in \citet{Bagdonaite:2014:10}, including
the improvements to the ThAr line-fitting and selection
\citep{Murphy:2007:221}, and the extraction of the ThAr flux using the
same spatial profiles weights as those established from the
corresponding quasar exposure. As described by
\citet{Malec:2010:1541}, those same improvements were implemented
within the {\sc
  hiredux}\footnote{{\urlstyle{rm}\url{http://www.ucolick.org/~xavier/HIRedux}}}
software suite which was used to reduce all the HIRES
exposures. Additional improvements to the optimal extraction and blaze
correction approach within {\sc hiredux} are also documented by
\citet{Malec:2016:thesis}.

The extracted spectra were then processed and co-added using {\sc
  uves\_popler} \citep{Murphy:2016:UVESpopler},
code specifically written for optimally combining pipeline-reduced
UVES and HIRES exposures. For UVES pipeline products, the pixel-space
flux and 1$\sigma$ error arrays from each echelle order, of each
exposure, are redispersed onto a common log-linear wavelength scale
after the wavelength solutions from the pipeline are converted to
vacuum in the heliocentric reference frame. {\sc hiredux} conducts
these steps itself and we have checked that they are consistent with
{\sc uves\_popler}'s implementation. {\sc uves\_popler} scales the
flux (and error) array in each echelle order to optimally match any
orders overlapping with it and then combines the flux values from all
exposures with inverse variance weighting in an iterative fashion to
remove `cosmic rays' and other artefacts. Any remaining artefacts were
removed manually upon inspection of the absorption features of
interest for this work. {\sc uves\_popler} also automatically fits a
low-order polynomial continuum which we inspected and adjusted with
local, low-order corrections around absorption features.

For each quasar, all extracted exposures from all runs on a single
spectrograph were initially combined to form a ``master spectrum'',
regardless of the differences in on-chip binning and spectral
resolution (i.e.\ slit width) between observing runs. Combining all
exposures allows for better artefact identification and a single
continuum fit to the highest-\SN\ representation of the
spectrum. However, the contributing exposures were then grouped
according to slit width and/or binning and saved as separate
``sub-spectra'' of the master. These sub-spectra are analysed
separately but simultaneously in our profile fitting approach, as
described in \Sref{ss:genfit}.

We make the final master spectrum or sub-spectra used in the fit for
each \daa\ measurement publicly available in
\citet{Murphy:2016:alphaZnCrII2016}.

\section{Analysis}\label{s:anal}

\subsection{Absorption profile fitting and \boldmath{$\daa$} measurement}\label{ss:genfit}

The \Ion{Zn/Cr}{ii} absorption profiles generally show a complex
velocity structure and, for each system, these are fitted with
multi-component Voigt profiles in order to derive the best-fitting
\daa\ which characterizes the pattern of any significant velocity
shifts between the transitions. Our approach to fitting the absorption
profiles is similar to many previous MM analyses
\citep[e.g.][]{Murphy:2001:1223,Murphy:2003:609,King:2012:3370,Molaro:2013:A68},
and the same as that described in detail by \citet{Evans:2014:128}. It
utilizes {\sc vpfit} version 9.5 \citep{Carswell:2014:VPFIT} to
minimize the $\chi^2$ between a model of the absorption profile and
the spectra. For a given absorber, the main steps and assumptions are
as follows:
\begin{itemize}
\item \daa\ is fixed to zero while a series of different models of the
  absorption profile is trialled, with successive models usually
  containing a larger number of fitted velocity components. The goal
  is to identify a single model that accounts for all the
  statistically significant structure in the observed profiles of the
  Zn and \Ion{Cr}{ii} transitions simultaneously.
\item In all trial models, the redshifts ($z$) and Doppler
  $b$-parameters of corresponding Zn and \Ion{Cr}{ii} components are
  tied together but their column densities, \Nion{Zn}{ii} and
  \Nion{Cr}{ii}, are not linked. The $b$-parameters are limited to
  $\ge$1\,\kms\ to avoid degeneracies with column densities in
  unresolved components and to ensure the $\chi^2$ minimization
  converges reasonably quickly.
\item The preferred model is the one with the smallest $\chi^2$ per
  degree of freedom, $\chi^2_\nu$. Other, similar criteria may be used
  to select the `best' model \citep[e.g.\ the Akaike information
  criterion,][]{Akaike:1974:716,King:2012:3370}, but the preferred
  model is usually either the same or very similar in these different
  approaches, so we use the simple $\chi^2_\nu$ criterion here, as in
  \citet{Murphy:2008:1053}.
\item \daa\ is only introduced as an additional free parameter once
  the preferred model is finalised. This prevents \daa\ from acquiring a spurious value during the process of establishing the preferred profile model. That process is also very interactive, so introducing \daa\ as a free parameter only afterwards avoids any human bias. \daa\ is then determined
  simultaneously with all the other free parameters (i.e.\ $z$, $b$,
  \Nion{Zn}{ii} and \Nion{Cr}{ii} for each component) via $\chi^2$
  minimization. The statistical uncertainty in \daa\ is derived from
  the appropriate diagonal term of the final covariance matrix;
  degeneracies with other fitted parameters are negligible.
\item We accept a fit, and the \daa\ resulting from it, into our statistical analysis of the
  results (\Sref{s:res}) only if 3 criteria are met by the preferred
  model: (i) $\chi^2_\nu<1.2$; (ii) no long-range ($\ga$5 pixel),
  significant and systematic excursions of the fit residuals in any
  transition; and (iii) no significant structure in the residuals
  common amongst the transitions, as judged from the `composite
  residual spectrum' (CRS) \citep{Malec:2010:1541} -- the residuals
  from all transitions normalized by their error spectra, registered
  on the same velocity axis and averaged. These criteria guard against
  cases where low-level artefacts in the spectra, that cannot be
  attributed to Gaussian noise, may cause an appreciable systematic
  effect in \daa. These selection criteria were applied after determining \daa\ to avoid rejecting fits in which poor residuals were caused by a real variation in $\alpha$. However, in practice, we observe no large differences between the residuals before and after determining \daa\ in any of the fits; none of the fits would have been rejected if the selection criteria were applied before \daa\ was determined.
\end{itemize}

Physically, the assumption implicit in tying together the
$b$-parameters of corresponding Zn and \Ion{Cr}{ii} components is that
turbulent broadening dominates over thermal broadening. The opposite
choice could equally have been made. However, there is little
practical difference between these assumptions for Zn and Cr because
of the small relative difference in their atomic masses
($\approx$23\,per cent; cf.\ $\approx$79\,per cent for Mg and
Fe). Furthermore, our aim is not to derive the physical properties of
individual components (if that is even possible in most systems) but
more practical: for determining \daa\ it is sufficient to obtain a
\emph{physically plausible fit where all the statistically significant
  structure in a system's absorption profile is accounted for by the
  preferred model}. As discussed in \citet{Evans:2014:128}, this
approach implies that, whatever broadening mechanism is assumed, the
differences they cause in the model are effectively marginalised over
in the $\chi^2$ minimization process and will not affect \daa\
significantly. The effect of different assumptions about the
broadening mechanisms on \daa\ has been explored by, e.g.,
\citet{Murphy:2003:609}, \citet{King:2012:3370} and
\citet{Evans:2014:128}, and found to be unimportant. Therefore, we fit
only turbulent velocity structures to our Zn and \Ion{Cr}{ii}
profiles.

A complication in fitting the Zn and \Ion{Cr}{ii} simultaneously, and
a difficulty when determining their (joint) velocity structure, is
that the weak \tran{Mg}{i}{2026} transition falls just 50\,\kms\
redwards of \tran{Zn}{ii}{2026}. In all but two of our absorbers
(those towards J0108$-$0037 and PHL957), the profile structure extends
over more than 50\,\kms\ and, because these are strong metal
absorbers, the \tran{Mg}{i}{2026} optical depth is sufficient to cause
significant blending with parts of the \tran{Zn}{ii}{2026}
profiles. This blend must be modelled carefully if \tran{Zn}{ii}{2026}
is to contribute meaningful, unbiased constraints on \daa. Our
approach is to fit the stronger -- i.e.\ much better defined --
\tran{Mg}{i}{2852} transition to strongly constrain the velocity
structure of the \tran{Mg}{i}{2026} blend. However, we completely
`decouple' the \Ion{Mg}{i} fit from the \Ion{Zn/Cr}{ii} fit in several
ways: (i) the fitted \Ion{Mg}{i} velocity structure need not match
that of the \Ion{Zn/Cr}{ii} fit in the same absorber (though, of
course, it will be similar); (ii) none of the \Ion{Mg}{i} component
redshifts or $b$-parameters are tied to any \Ion{Zn/Cr}{ii}
components; and (iii) a velocity shift between the two \Ion{Mg}{i}
transitions is fitted as an additional free parameter. This decoupling
ensures that long-range systematic effects that generate spurious
shifts between the \tran{Mg}{i}{2026} and 2852 transitions do not
affect the \daa\ parameter in the fit. We find the best-fitted velocity shifts to be typically $\sim$100--600\,\ms, as expected for long-range distortions of the magnitude found in previous studies \citep[i.e.\ $\la$0.3\,\ms\,\AA$^{-1}$, e.g.][]{Whitmore:2015:446}.

In cases where the master spectrum was split into two or more
``sub-spectra'', with different resolutions (slit width) and/or
dispersion (CCD binning), the sub-spectra were fitted simultaneously
with the same absorption profile model. To allow for systematic shifts
of sub-spectra in these different instrument and/or observing set-ups,
an additional velocity shift between them was fitted as a free
parameter. Again, this serves to decouple such systematics from the
\daa\ parameter in the fit. We find these shifts to have a magnitude $<$300\,\ms, similar to the typical values found in other studies \citep[e.g.][]{Evans:2014:128}.

The laboratory wavelengths and oscillator strengths used for this
analysis were reviewed in \citet{Murphy:2014:388}. These include
isotopic structures calculated for the \Ion{Cr}{ii} transitions
\citep{Berengut:2011:052520} and \tran{Zn}{ii}{2062}
\citep{Berengut:2003:022502,Matsubara:2003:209}, the measured isotopic
structure of \tran{Zn}{ii}{2026} \citep{Matsubara:2003:209}, the
calculated hyperfine structure for the transitions of
$^{67}$\Ion{Zn}{ii} \citep{Campbell:1997:2351,Dixit:2008:025001}, and
the measured isotopic component wavelengths for \tran{Mg}{i}{2026}
\citep{Hannemann:2006:012505} and \tran{Mg}{i}{2852}
\citep{Salumbides:2006:L41}. The laboratory wavelengths all have
uncertainties $<$10\,\ms; errors in these wavelengths should cause
negligible systematic errors in \daa.

\subsection{Fits to individual absorption systems}\label{ss:fits}

Below we discuss the fit to each absorption system, highlighting
complexities or difficulties where relevant. The fitted parameters and
their uncertainties are provided in electronic, {\sc vpfit} format in
\citet{Murphy:2016:alphaZnCrII2016} for full transparency and
reproducibility. \Tref{t:res} summarizes the main results from each
fit, including the best-fitted value of \daa, its statistical
uncertainty and $\chi^2_\nu$.

\begin{table}
\begin{center}
  \caption{Main result for each absorber. The quasar name, absorption
    redshift (\zab) and telescope (Tel.) identify the absorbers
    plotted in Figs.\ \ref{f:J0058H}--\ref{f:Q2206}. The best-fitted
    value of \daa\ and its 1$\sigma$ statistical uncertainty
    ($\sigma_{\rm stat}$) from the $\chi^2$ minimization process are
    provided, along with the final $\chi^2$ per degree of freedom,
    $\chi^2_\nu$. The systematic uncertainty estimate for each
    absorber ($\sigma_{\rm sys}$) is the quadrature sum of
    uncertainties from long-range distortions (0.3\,ppm per absorber;
    \Sref{sss:long}), intra-order distortions (2.03\,ppm per absorber;
    \Sref{sss:intra}), and redispersion effects (0.49--2.28\,ppm;
    \Sref{sss:redisp}). These results were derived using the terrestrial isotopic abundances for Zn and Cr; see discussion of this potential systematic error in \Sref{sss:iso}. Note that the result from the VLT spectrum of
    J0058$+$0041, shown in italics, fails our selection criteria
    (\Sref{ss:genfit}) and is not considered in our statistical
    analysis in \Sref{ss:mainres}.}
\label{t:res}
\begin{tabular}{:l;c;l;c;c;c;c}\hline 
Quasar & \zab & Tel. & \daa & $\sigma_{\rm stat}$ & $\sigma_{\rm sys}$ & $\chi^2_\nu$ \\
& & & [ppm] & [ppm] & [ppm] & \\\hline
J0058$+$0041 & 1.072 & Keck & $ -1.35 $ &  6.71 &  2.51 & 0.91 \\ 
\rowstyle{\itshape}J0058$+$0041 & 1.072 & VLT  & { 17.07 } &  9.00 &  2.41 & 1.43 \\ 
PHL957       & 2.309 & Keck & $ -0.65 $ &  6.46 &  2.26 & 0.99 \\ 
PHL957       & 2.309 & VLT  & $ -0.20 $ & 12.44 &  3.51 & 1.09 \\ 
J0108$-$0037 & 1.371 & VLT  & $ -8.45 $ &  5.69 &  4.64 & 1.08 \\ 
J0226$-$2857 & 1.023 & VLT  & $  3.54 $ &  8.54 &  2.38 & 1.02 \\ 
J0841$+$0312 & 1.342 & Keck & $  3.05 $ &  3.30 &  2.13 & 0.77 \\ 
J0841$+$0312 & 1.342 & VLT  & $  5.67 $ &  4.19 &  2.16 & 1.04 \\ 
J1029$+$1039 & 1.622 & Keck & $ -1.70 $ &  9.80 &  2.47 & 0.70 \\ 
J1237$+$0106 & 1.305 & Keck & $ -4.54 $ &  8.08 &  3.13 & 0.76 \\ 
Q1755$+$57   & 1.971 & Keck & $  4.72 $ &  4.18 &  2.16 & 0.86 \\ 
Q2206$-$1958 & 1.921 & VLT  & $ -4.65 $ &  6.01 &  2.24 & 1.06 \\ 
\hline
\end{tabular}

\end{center}
\end{table}

\subsubsection{$\zab=1.072$ towards J0058$+$0041}\label{sss:J0058}

\begin{figure}
\begin{center}
\includegraphics[width=0.99\columnwidth]{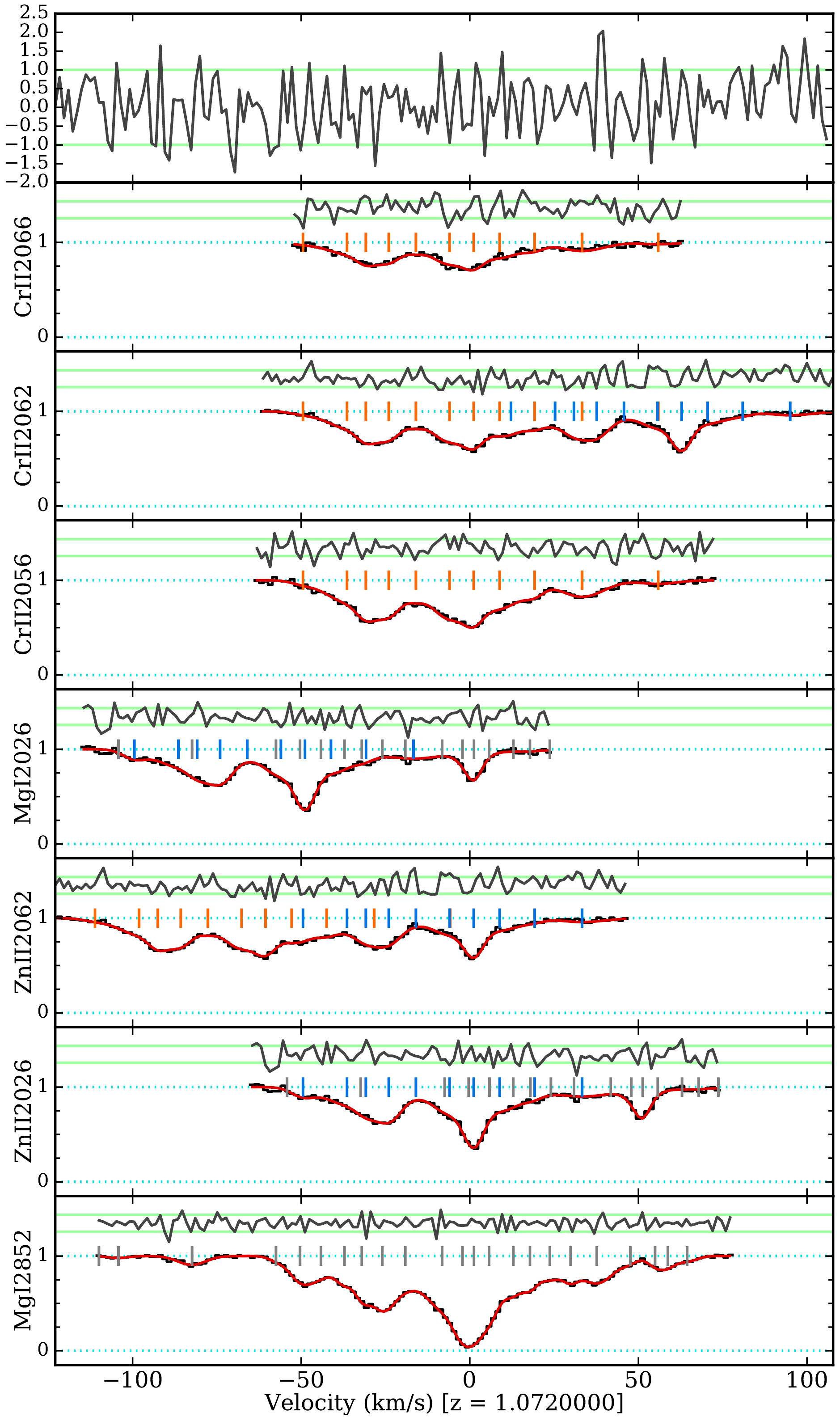}\vspace{-1em}
\caption{Fit to the Keck/HIRES spectrum of the $\zab=1.072$ absorber
  towards J0058$+$0041. The lower seven panels each show the
  continuum-normalized flux versus velocity (relative to the redshift
  in brackets) for one of the \Ion{Zn}{ii}, \Ion{Cr}{ii} and/or
  \Ion{Mg}{i} transitions fitted to measure \daa. The normalized
  spectrum (black histogram) is overlaid by the fit (red solid line)
  which comprises Voigt profile components at the velocities marked by
  the ticks (short vertical lines; orange for \Ion{Cr}{ii}, blue for
  \Ion{Zn}{ii} and grey for \Ion{Mg}{i}). The residual spectrum,
  normalized by the 1$\sigma$ error spectrum, is plotted (grey line)
  above the spectrum, with its $\pm$1$\sigma$ levels marked by the
  green solid lines. The top panel shows the composite residual
  spectrum (CRS; grey line) in units of standard deviations, relative
  to its $\pm$1$\sigma$ levels (green solid lines), derived by
  stacking the normalized residuals of all transitions shown. Note
  that the \tran{Zn}{ii}{2062} and \tran{Cr}{ii}{2062} transitions are
  blended with each other, as are \tran{Zn}{ii}{2026} and
  \tran{Mg}{i}{2026}; our approach to fitting these blends, using the
  \tran{Mg}{i}{2852} line, is discussed in \Sref{ss:genfit}.}
\label{f:J0058H}
\end{center}
\end{figure}

\begin{figure*}
\begin{center}
\includegraphics[width=0.75\textwidth]{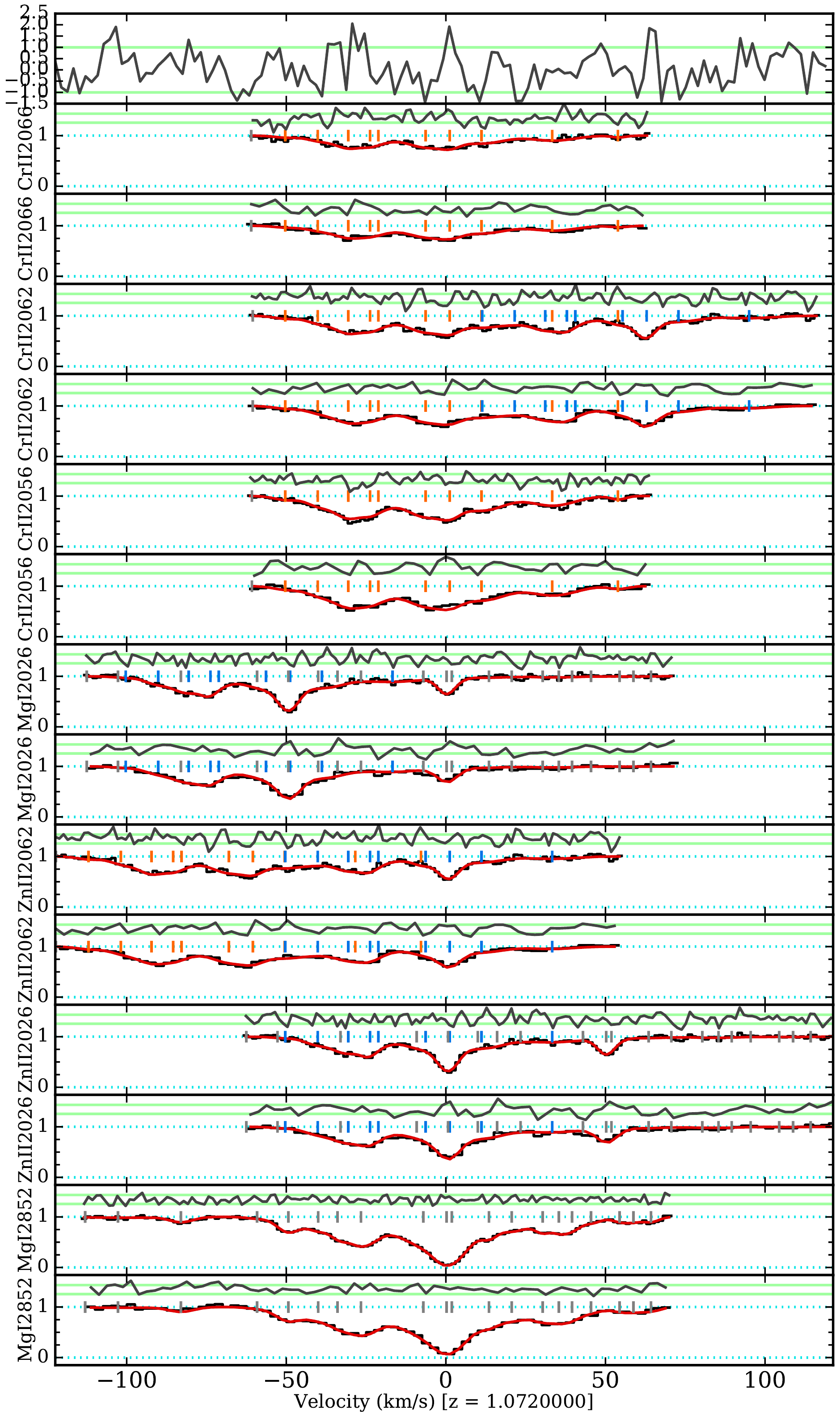}\vspace{-1em}
\caption{Same as \Fref{f:J0058H} but for the VLT/UVES spectrum of the
  $\zab=1.072$ absorber towards J0058$+$0041. Two sub-spectra are
  shown for each transition, one comprising unbinned exposures (upper
  panel for each transition) and another comprising 2$\times$2-binned
  exposures (lower panel for each transition). Note that these data
  and the fit fails our selection criteria for inclusion in our
  main results (see text).}
\label{f:J0058U}
\end{center}
\end{figure*}

This absorber extends over 120\,\kms\ in \Ion{Zn/Cr}{ii} but has 2
main (i.e.\ stronger and better defined) spectral features at $-30$ and
0\,\kms, as shown in \Fref{f:J0058H} (the fit to the HIRES spectrum)
and \Fref{f:J0058U} (UVES). These contribute the main constraints on
\daa. The blending \tran{Mg}{i}{2026} velocity structure extends over
much of the \tran{Zn}{ii}{2026} transition but, as explained in
\Sref{ss:genfit} above, \tran{Mg}{i}{2852} is used in both the HIRES
and UVES spectra to model this blend (visually, this may be more easily appreciated by first referring to the simpler, narrower absorption towards J0108$-$0037 in \Fref{f:J0108}). The HIRES and UVES profiles were
fit separately, with similar but not identical models, providing
independent \daa\ measurements. The UVES spectrum comprised two
sub-spectra, composed of the unbinned and 2$\times$2-binned exposures,
and these were fitted simultaneously as separate spectra (i.e.\ with the
same absorption profile model) with a velocity shift between them as
an additional free parameter, as described in \Sref{ss:genfit}. Visual
inspection of \Fref{f:J0058U} shows no evidence of substantial
inconsistencies between the two differently binned spectra.

Two small differences between the HIRES and UVES models in Figs.\
\ref{f:J0058H} \& \ref{f:J0058U} are worth considering as examples:
\begin{enumerate}
\item One velocity component, at $-20$\,\kms\ in the HIRES model, is
  fitted at slightly lower velocity in the UVES model. The weak,
  broader absorption this component is responsible for means that it
  will have no significant effect on \daa;
\item The reddest component in \Ion{Cr}{ii} is too weak to be fitted
  in \Ion{Zn}{ii}. It is present in both the HIRES and UVES models
  but, given the \SN\ of the spectra, it was statistically unnecessary
  to fit in the \Ion{Zn}{ii} transitions.
\end{enumerate}
The effect of such weak, often broad velocity components on \daa\ is
negligible, so we omit discussion of similar cases in the rest of this
section, focussing instead on components which, at first, may be
suspected of having affected \daa\ more significantly.

The fit to the HIRES spectrum in \Fref{f:J0058H} shows no evidence of
significant, many-pixel excursions in the residuals of any transition,
nor in the composite residual spectrum (CRS), and the final
$\chi^2_\nu=0.91$. Therefore, the best-fitted \daa\ value shown in
\Tref{t:res}, $-1.4\pm6.7_{\rm stat}$, is accepted for our main results.

By contrast, the fit to the UVES spectrum in \Fref{f:J0058U} shows a
run of significant residuals at $-27$\,\kms\ in \tran{Cr}{ii}{2056} in
the unbinned spectrum, the CRS may have some significant structures,
and \Tref{t:res} shows a $\chi^2_\nu=1.43$. An extensive set of trial
profile models was explored to remedy these problems, but even models
with significantly more velocity components did not solve them. We
therefore reject this system from our main results in
\Sref{s:res}. This is the only one of the 12 possible \daa\
measurements that we reject in this way.

\subsubsection{$\zab=2.309$ towards PHL957}\label{sss:PHL957}

\begin{figure}
\begin{center}
\includegraphics[width=0.99\columnwidth]{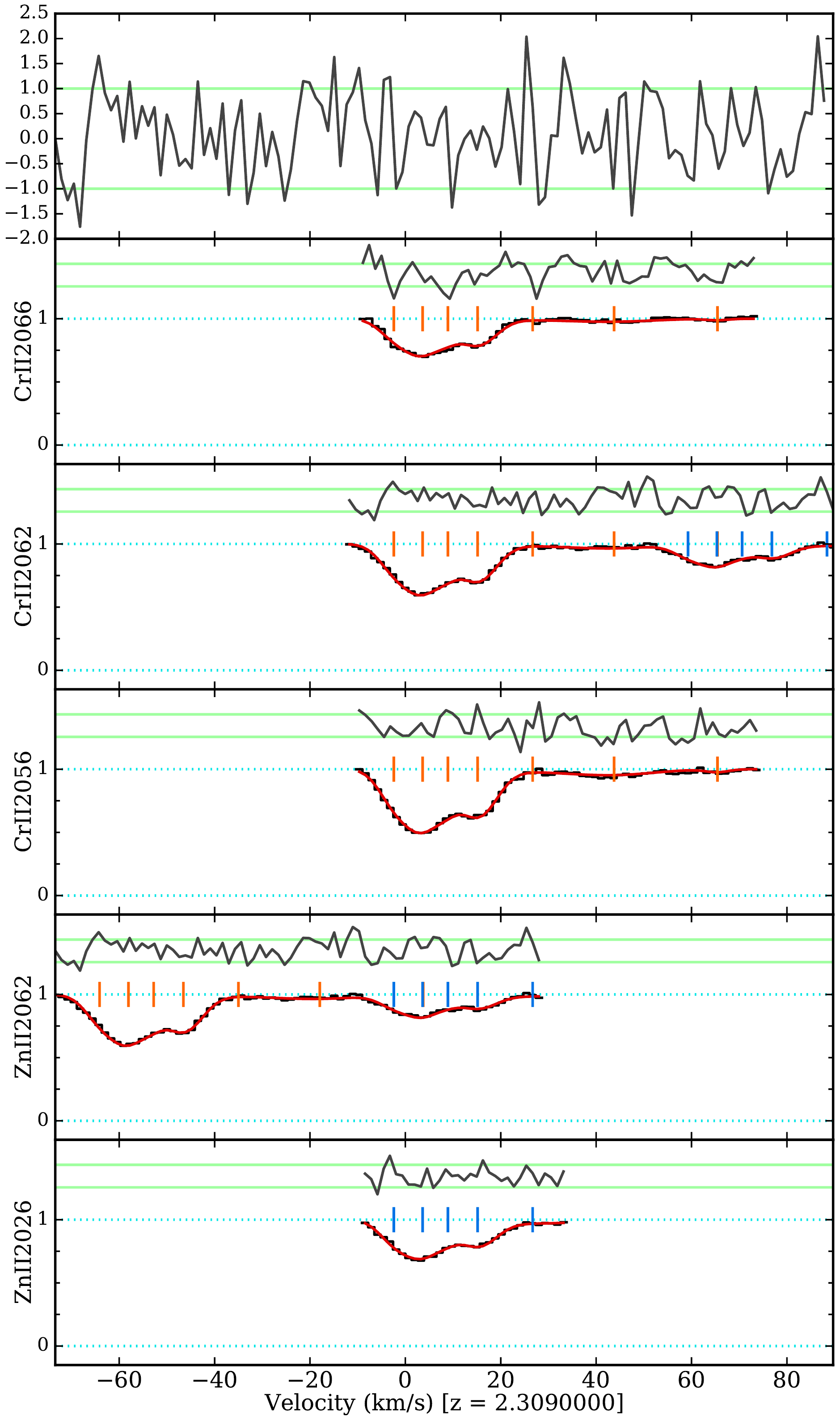}\vspace{-1em}
\caption{Same as \Fref{f:J0058H} but for the Keck/HIRES spectrum of
  the $\zab=2.039$ absorber towards PHL957. \tran{Mg}{i}{2852} is
  heavily blended with telluric absorption but is not required because
  the narrow absorption profile means \tran{Mg}{i}{2026} does not
  blend with \tran{Zn}{ii}{2026}.}
\label{f:PHL957H}
\end{center}
\end{figure}

\begin{figure}
\begin{center}
\includegraphics[width=0.99\columnwidth]{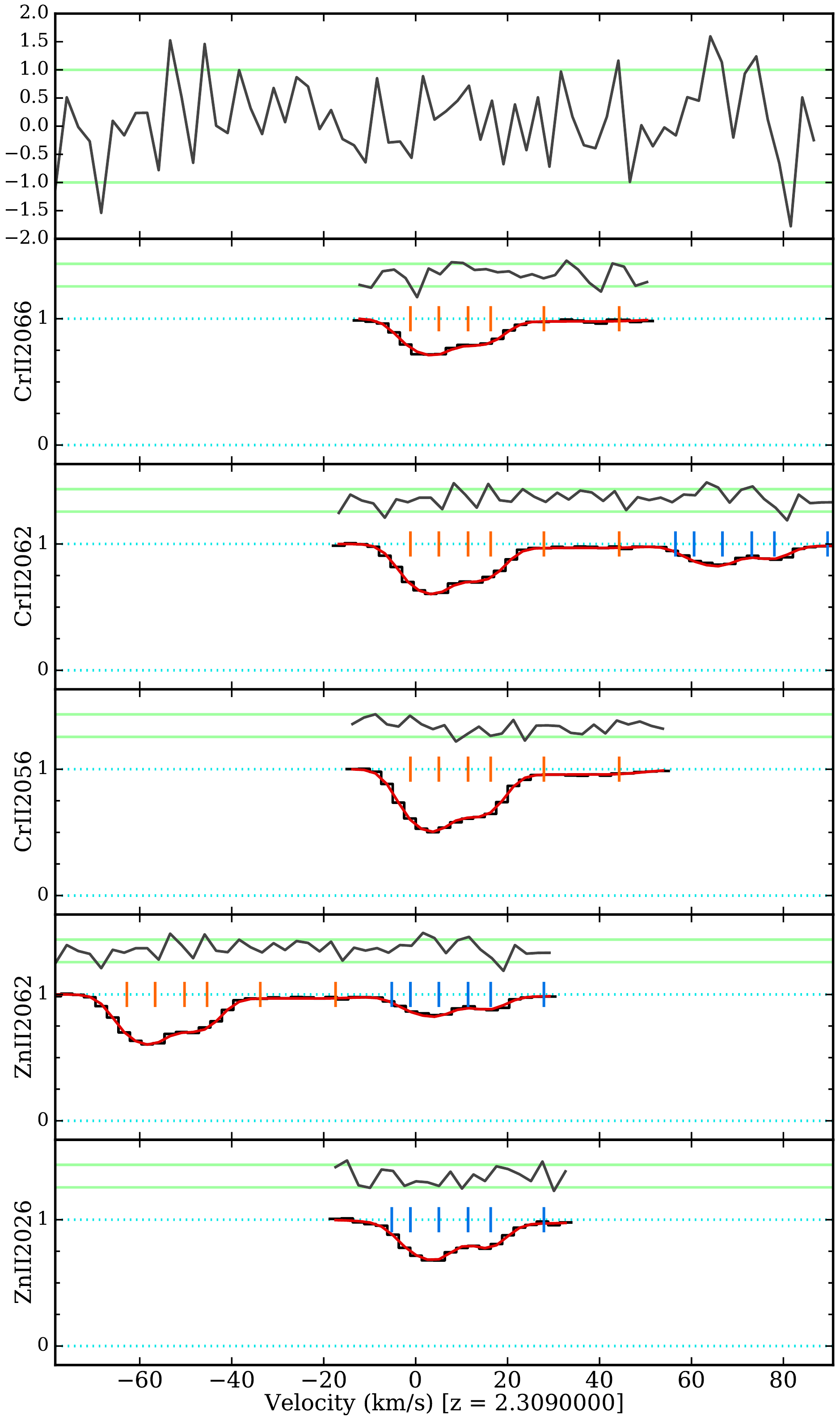}\vspace{-1em}
\caption{Same as \Fref{f:PHL957H} but for the VLT/UVES spectrum of the
  $\zab=2.309$ absorber towards PHL957.}
\label{f:PHL957U}
\end{center}
\end{figure}

Figures \ref{f:PHL957H} and \ref{f:PHL957U} show our fits to the HIRES
and UVES spectra of this absorber, which is relatively narrow,
comprising two main spectral features spanning just
$\approx$40\,\kms. Therefore, there is no blending from
\tran{Mg}{i}{2026} in these main features of \tran{Zn}{ii}{2026}. This
is important because, at this redshift, the \tran{Mg}{i}{2852} is
heavily blended with telluric absorption and so cannot be used to help
constrain any \tran{Mg}{i}{2026} absorption. We detect very weak
absorption at 30--70\,\kms\ in the \Ion{Cr}{ii} transitions. Much
stronger transitions (e.g.\ \tran{Al}{ii}{1670}, \tran{Si}{ii}{1526},
\tran{Fe}{ii}{2374} etc.) show another two spectral features at these
velocities, so our fit includes two weak velocity components to
account for them. However, we do not include those components in the
\Ion{Zn}{ii} fit because they are so weak, providing no significant
constraints on \daa\ and, in \tran{Zn}{ii}{2026}, they would be
blended with any weak \tran{Mg}{i}{2026} absorption.

The fits to both the HIRES and UVES spectra in Figs.\ \ref{f:PHL957H}
and \ref{f:PHL957U} show no evidence of significant, many-pixel
excursions in the residuals of any transition, nor in the CRSs, and
the final $\chi^2_\nu$ values in \Tref{t:res} satisfy our selection
criterion in \Sref{ss:genfit}. Therefore, both results from this
absorber are accepted in our main results.

\subsubsection{$\zab=1.371$ towards J0108$-$0037}\label{sss:J0108}

\begin{figure}
\begin{center}
\includegraphics[width=0.99\columnwidth]{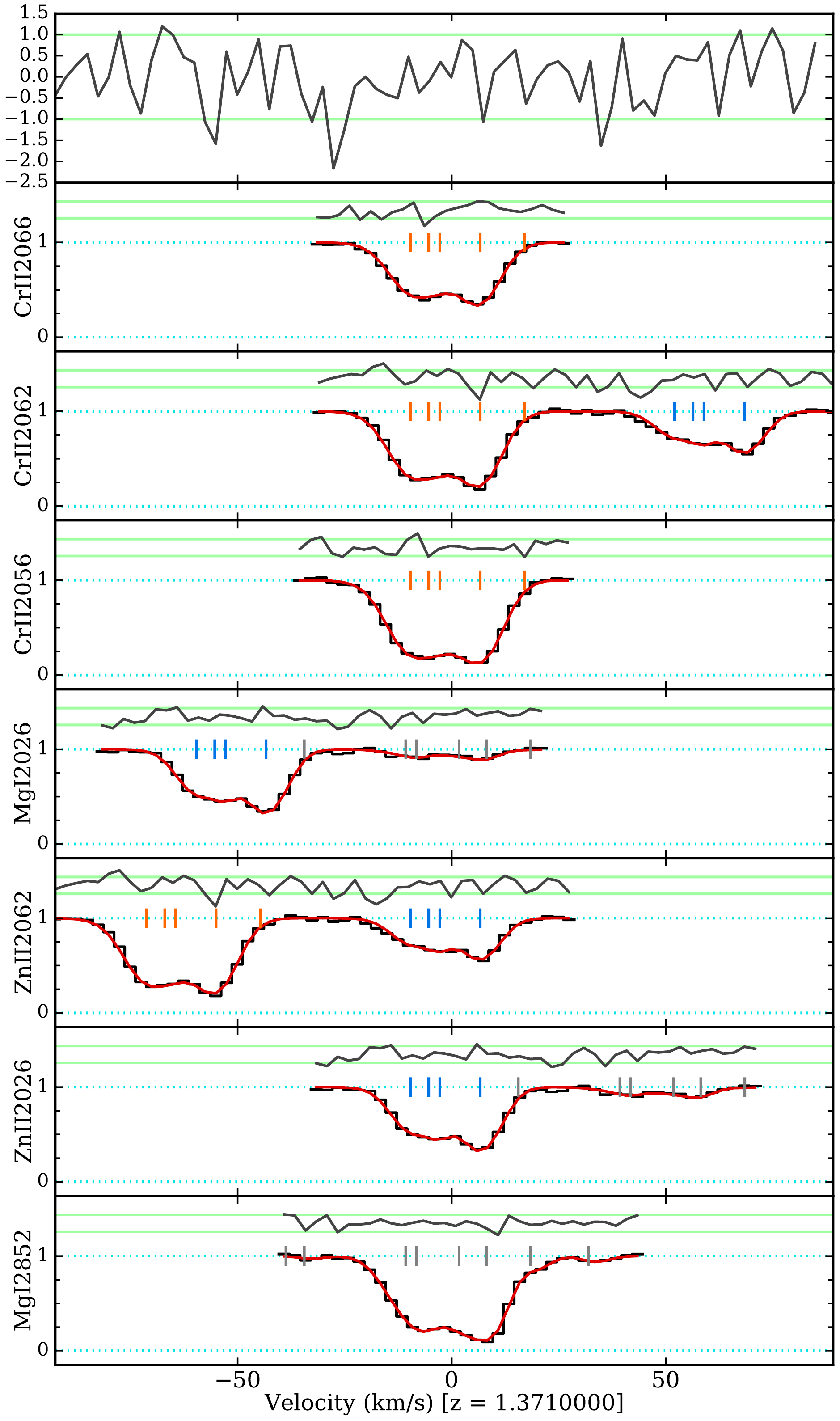}\vspace{-1em}
\caption{Same as \Fref{f:J0058H} but for the VLT/UVES spectrum of the $\zab=1.371$ absorber towards J0108$-$0037.}
\label{f:J0108}
\end{center}
\end{figure}

\Fref{f:J0108} shows that the Zn and \Ion{Cr}{ii} absorption in this
system has a very similar structure to that in the absorber towards
PHL957, with two main features spanning only $\approx$40\,\kms. Three
closely-spaced components are required in the bluer spectral feature,
with one component (the reddest of the three) being very narrow, just
above our $b=1.0$-\kms\ limit. A fit with only 2 components in this
spectral feature has substantially poorer $\chi^2_\nu$ so is rejected
in preference to the 3-component model. The lower redshift of this
system allows \tran{Mg}{i}{2852} to be included, and \Fref{f:J0108}
shows that no blending of \tran{Mg}{i}{2026} with \tran{Zn}{ii}{2026}
occurs. Indeed, as was necessary for PHL957, the \Ion{Mg}{i} fit could
have been entirely neglected from this absorber and the fit to
\tran{Zn}{ii}{2026} simply truncated at $\approx$25\,\kms. We retain
these features here to demonstrate these points for PHL957, but they
have no bearing on \daa. \tran{Mg}{i}{2852} also shows that a weak
component exists in the red wing of the redder spectral feature, and
we find that a similar component is required to fit the \Ion{Cr}{ii}
lines. However, this component is very weak, and is not required to
fit the weaker \Ion{Zn}{ii} transitions.

The fit to the UVES spectrum in \Fref{f:J0108} satisfies our selection
criteria in \Sref{ss:genfit}, so is accepted in our main results.

\subsubsection{$\zab=1.023$ towards J0226$-$2857}\label{J0226}

\begin{figure*}
\begin{center}
\includegraphics[width=0.75\textwidth]{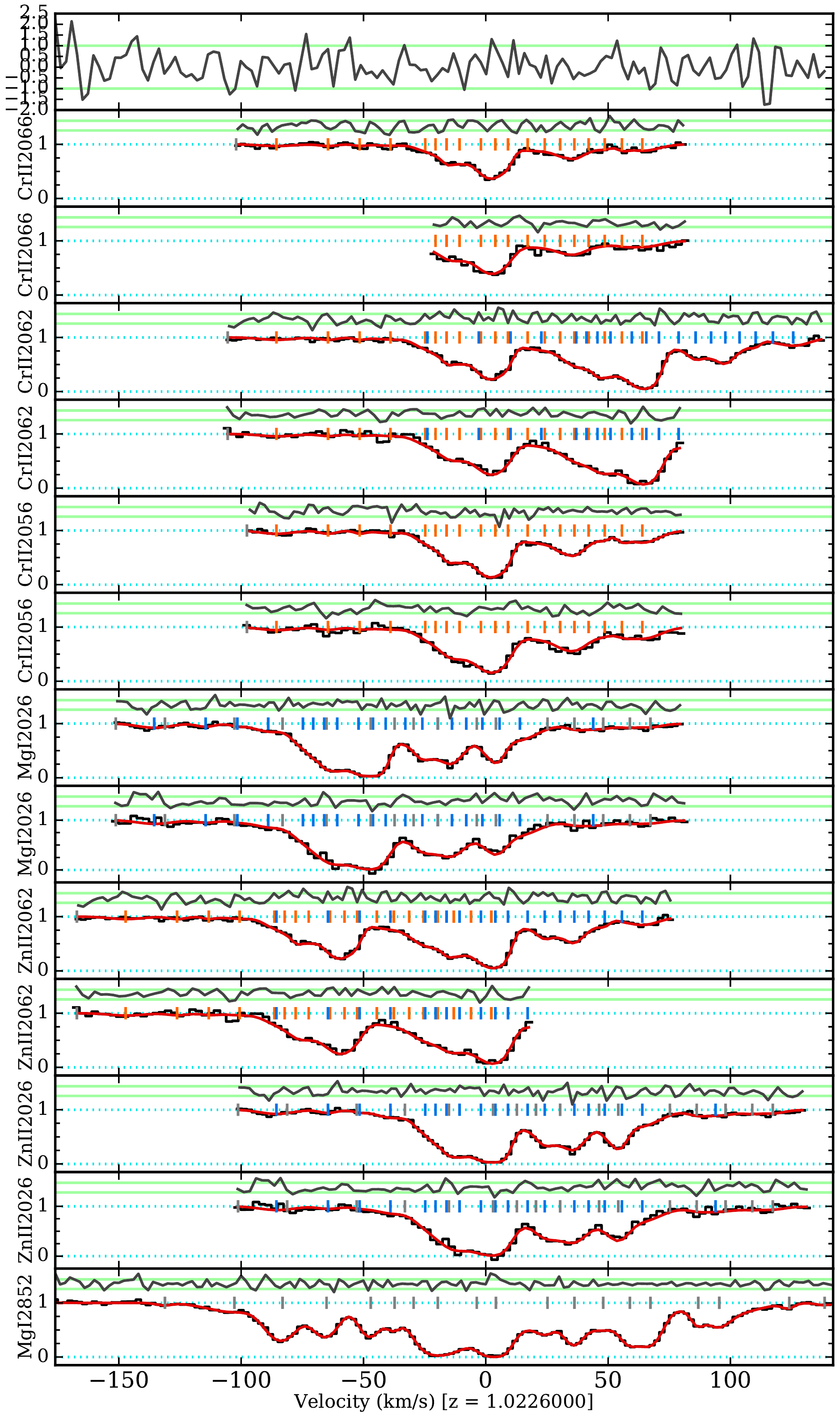}\vspace{-1em}
\caption{Same as \Fref{f:J0058U} but for the VLT/UVES spectrum of the
  $\zab=1.023$ absorber towards J0226$-$2857. Two sub-spectra are
  shown for each transition, one comprising higher-resolution,
  0.8\,arcsec slit-width exposures with 2.0\,\kms\ dispersion (upper
  panel for each transition) and another comprising lower-resolution,
  1.2\,arcsec slit-width exposures with 2.5\,\kms\ dispersion (lower
  panel for each transition). The \tran{Mg}{i}{2852} absorption
  continues at velocities $>$140\,\kms\ but has no bearing on the
  blending of \tran{Mg}{i}{2026} with \tran{Zn}{ii}{2026} and is not
  shown.}
\label{f:J0226}
\end{center}
\end{figure*}

Two sub-spectra were created from the master UVES spectrum for
analysis of this absorber, one from the 4 higher-resolution,
0.8\,arcsec slit-width exposures, the other from the 3
lower-resolution, 1.2\,arcsec slit-width exposures (see
\Tref{t:obs}). All exposures had 2$\times$2 on-chip CCD binning
(native bin size is $\approx$2.3\,\kms) so, to reduce information loss
from redispersion when combining the exposures, we used two different
dispersions, 2.0 and 2.5\,\kms, for the higher and lower-resolution
sub-spectra, respectively.

\Fref{f:J0226} shows the separate, simultaneous fit to both
sub-spectra, with a velocity shift between them fitted as an
additional free parameter, as described in \Sref{ss:genfit}. The
velocity structure is quite complex, spanning 180\,\kms\ with 3 main,
fairly separate spectral features in Zn and \Ion{Cr}{ii}. The bluest
two of these \Ion{Zn/Cr}{ii} features must be fitted with 6--7
velocity components each. The most important part of the profile for
constraining \daa\ is the deepest, sharpest feature at 0\,\kms,
especially its sharp red wing. The other features, being smoother and
shallower, constrain \daa\ correspondingly less well. No substantial
differences were required between the velocity structures fitted to
the Zn and \Ion{Cr}{ii} transitions. With such a broad velocity
structure, blending with \tran{Mg}{i}{2026} is important for the
\tran{Zn}{ii}{2026} fit. For this purpose, the strong, very broad
(380\,\kms\ with $\approx$10 spectral features) profile of
\tran{Mg}{i}{2852} was available in the higher-resolution sub-spectrum
at a high \SN.

Preliminary fits revealed that the lower-resolution sub-spectrum
contained two small ($\approx$6-pixel) regions of significantly lower
flux compared to the high-resolution sub-spectrum: in
\tran{Zn}{ii}{2062} at 25\,\kms\ and in \tran{Cr}{ii}{2066} at
$-30$\,\kms. We therefore truncated the blue and red edges of the
fitting regions in these transitions, in the lower-resolution
sub-spectrum only, just above and below these velocities,
respectively. Including these regions in our fiducial fit does not
change \daa\ significantly (it decreases by $\approx$0.6\,ppm, less
than a tenth of the statistical error in this absorber) but would
disqualify this absorber from entering our statistical sample. Indeed,
comparing sub-spectra in this way is an effective means of checking
for artefacts that may otherwise cause systematic effects in
individual systems.

The final fit to the two UVES sub-spectra in \Fref{f:J0226} shows no
evidence of significant problems with the residuals in individual
transitions or the CRS, and the final $\chi^2_\nu$ in \Tref{t:res}
satisfies our criterion in \Sref{ss:genfit} for accepting this
absorber into our main results.

\subsubsection{$\zab=1.342$ towards J0841$+$0312}\label{sss:J0841}

\begin{figure}
\begin{center}
\includegraphics[width=0.99\columnwidth]{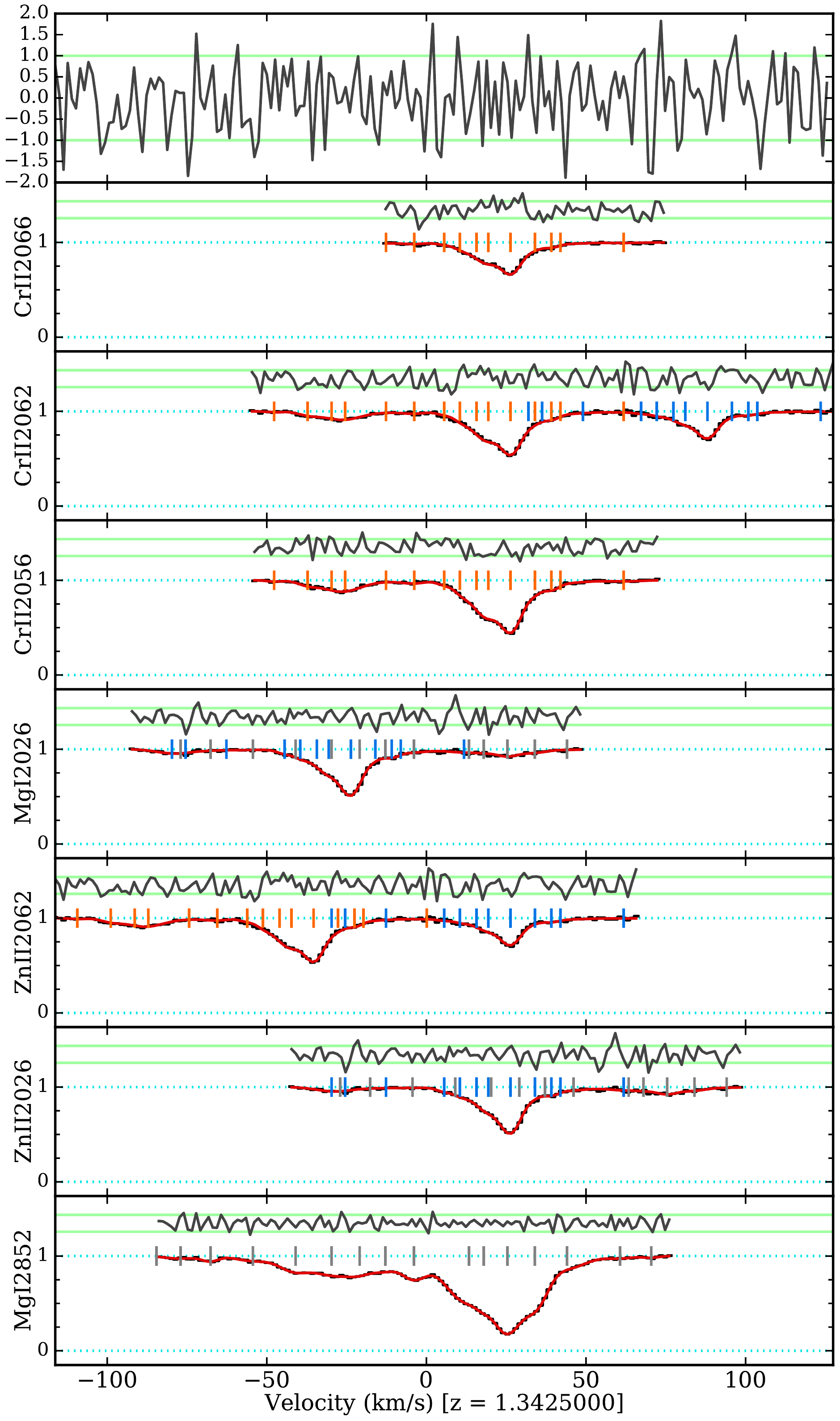}\vspace{-1em}
\caption{Same as \Fref{f:J0058H} but for the Keck/HIRES spectrum of
  the $\zab=1.342$ absorber towards J0841$+$0312.}
\label{f:J0841H}
\end{center}
\end{figure}

\begin{figure}
\begin{center}
\includegraphics[width=0.99\columnwidth]{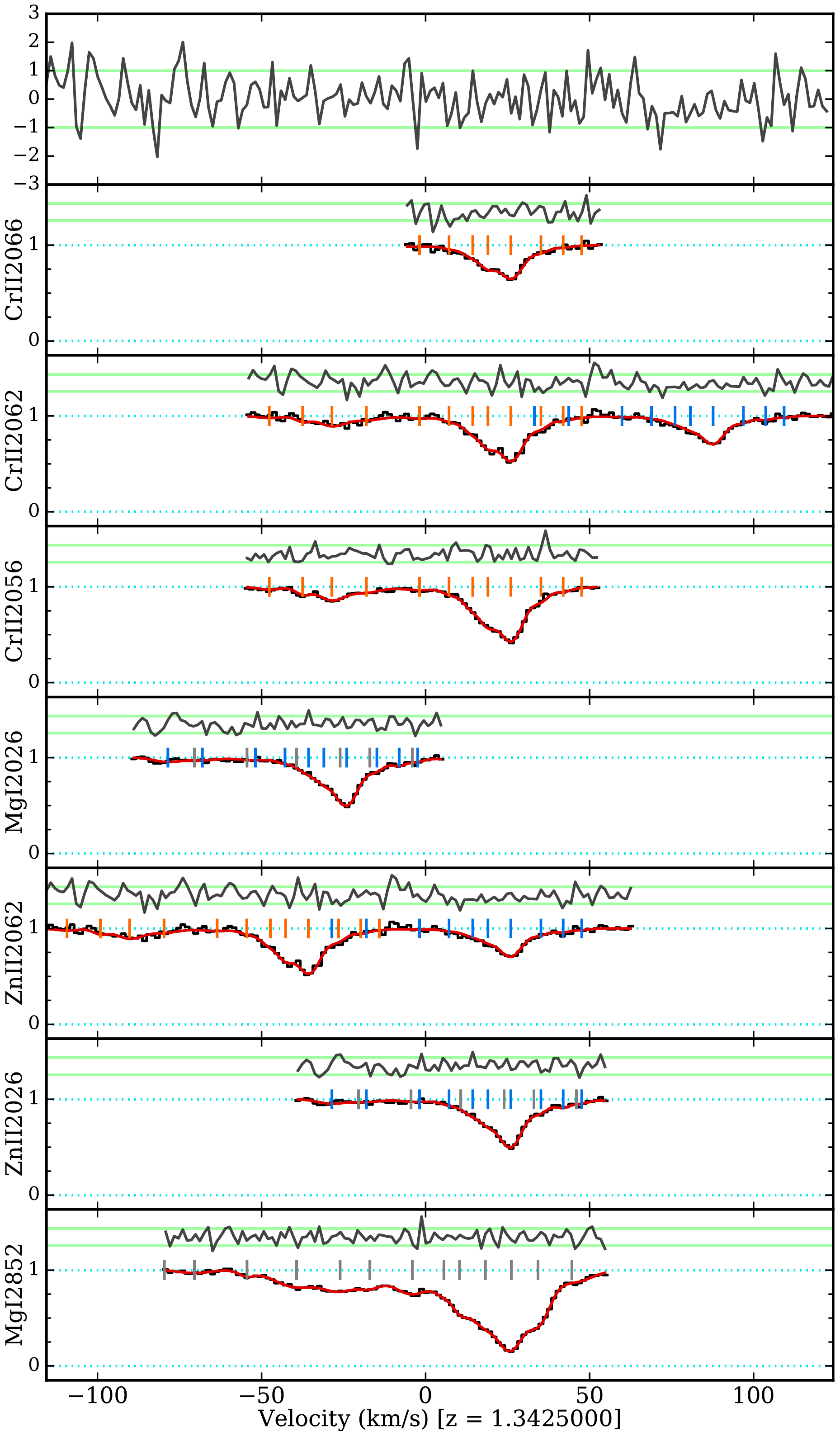}\vspace{-1em}
\caption{Same as \Fref{f:J0841H} but for the VLT/UVES spectrum of the
  $\zab=1.342$ absorber towards J0841$+$0312.}
\label{f:J0841U}
\end{center}
\end{figure}

Figures \ref{f:J0841H} and \ref{f:J0841U} show our fits to the HIRES
and UVES spectra of this absorber. \Ion{Zn/Cr}{ii} absorption is
detected over 120\,\kms\ in two spectral features, with the deepest
one around 25\,\kms\ constraining \daa\ most strongly. The relative
weakness of the spectral feature at $-80$\,\kms\ means that the
\tran{Mg}{i}{2026} blending with the main spectral feature of
\tran{Zn}{ii}{2026} is very weak (though not entirely
negligible). There is additional absorption from a blend (from another
redshift) in the blue spectral feature of \tran{Cr}{ii}{2066}, below
$\approx-15$\,\kms, so this spectral feature is not included in the
fit to that transition.

The fits to both the HIRES and UVES spectra in Figs.\ \ref{f:J0841H}
and \ref{f:J0841U} show no evidence problems in the residuals, nor in
the CRSs, and the final $\chi^2_\nu$ values in \Tref{t:res} satisfy
our selection criterion in \Sref{ss:genfit} for accepting both results
for this absorber into our main results.

\subsubsection{$\zab=1.622$ towards J1029$+$1039}\label{sss:J1029}

\begin{figure}
\begin{center}
\includegraphics[width=0.99\columnwidth]{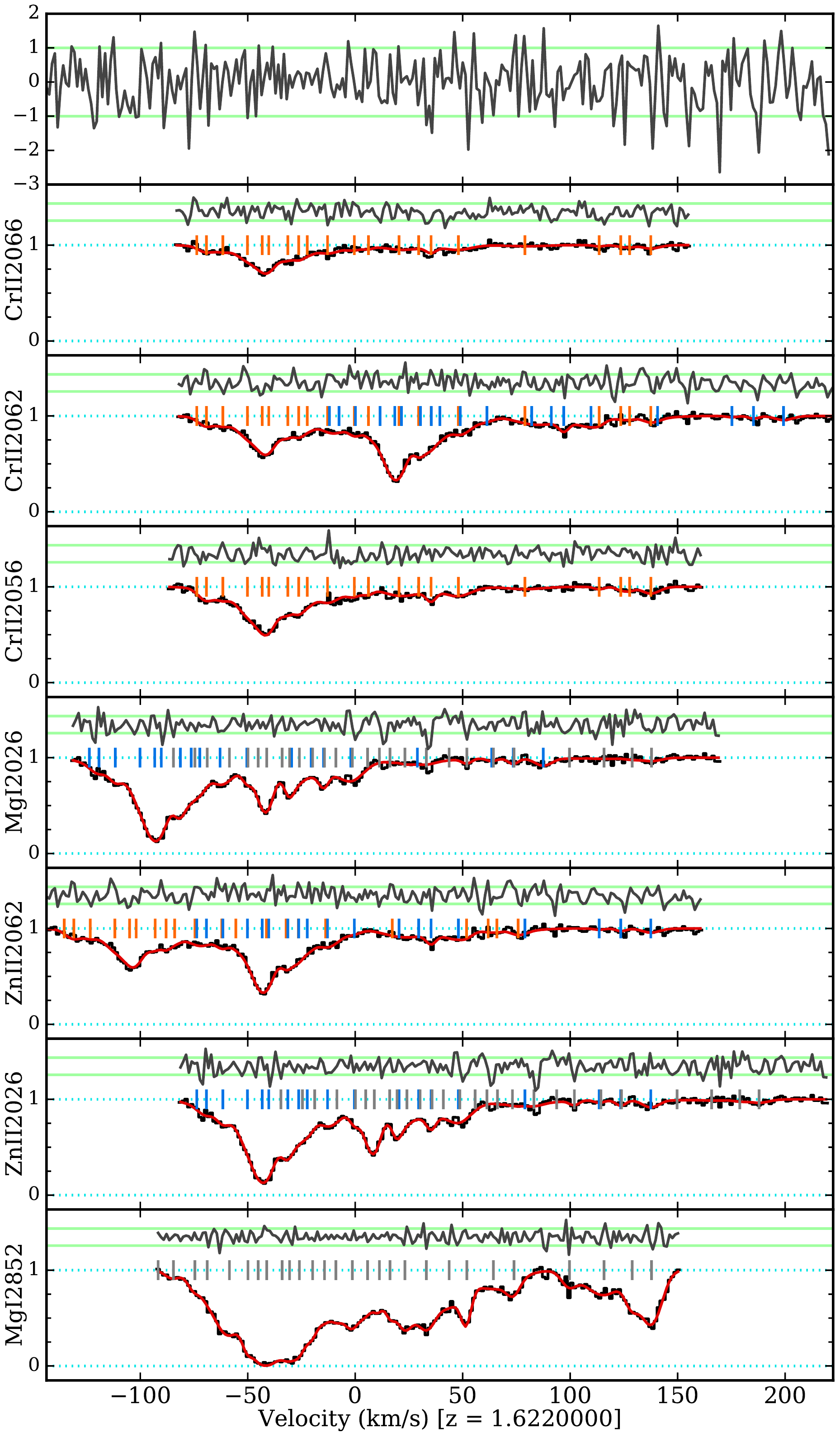}\vspace{-1em}
\caption{Same as \Fref{f:J0058H} but for the Keck/HIRES spectrum of
  the $\zab=1.622$ absorber towards J1029$+$1039.}
\label{f:J1029}
\end{center}
\end{figure}

The Keck/HIRES spectrum of this absorber in \Fref{f:J1029} shows very
complex and broad velocity structure, spanning 230\,\kms\ in
\Ion{Zn/Cr}{ii}, dominated by one main spectral feature at
$-40$\,\kms\ comprising many velocity components. The overall shape of
that feature is relatively broad and smooth, with few sharp
sub-features, leading to a weaker constraint on \daa\ than in most
other absorbers studied here. No substantial differences were required
between the velocity structures fitted to the Zn and \Ion{Cr}{ii}
transitions. With such a broad velocity structure, blending with
\tran{Mg}{i}{2026} is important for the \tran{Zn}{ii}{2026} fit; the
high \SN\ of the \tran{Mg}{i}{2852} transition is more than adequate
for this purpose. \Fref{f:J1029} shows no evidence of significant,
many-pixel deviations in the residuals in individual transitions or
the CRS, and the final $\chi^2_\nu$ in \Tref{t:res} satisfies our
criterion in \Sref{ss:genfit} for accepting this absorber into our
main results.

\subsubsection{$\zab=1.305$ towards J1237$+$0106}\label{sss:J1237}

\begin{figure}
\begin{center}
\includegraphics[width=0.99\columnwidth]{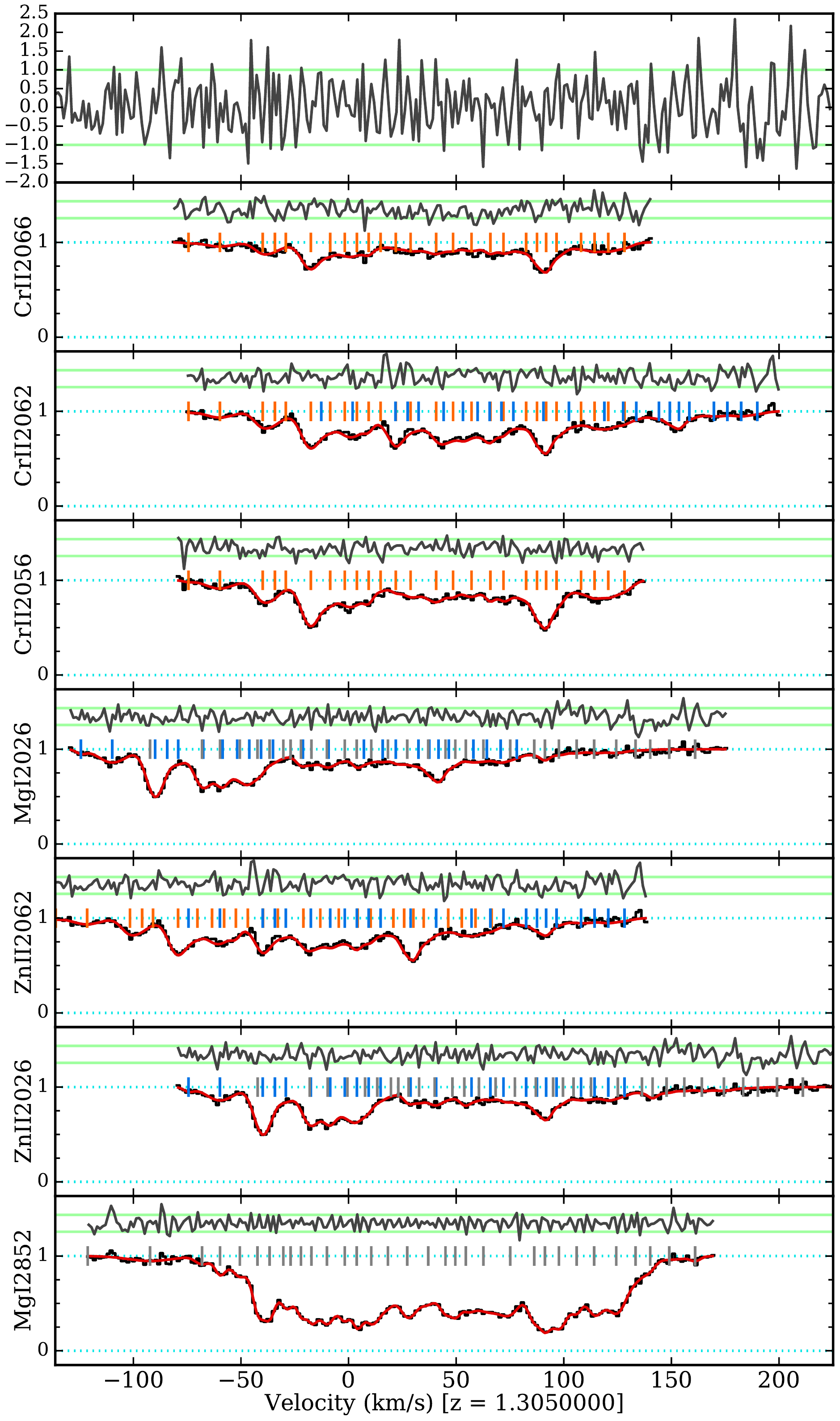}\vspace{-1em}
\caption{Same as \Fref{f:J0058H} but for the Keck/HIRES spectrum of
  the $\zab=1.305$ absorber towards J1237$+$0106.}
\label{f:J1237}
\end{center}
\end{figure}

The Keck/HIRES spectrum of this absorber in \Fref{f:J1237} shows many
similarities to that of J1029$+$1039 in \Fref{f:J1029}, with very
complex and broad velocity structure (spanning 215\,\kms\ in
\Ion{Zn/Cr}{ii}). In this case the constraints on \daa\ are dominated
by two stronger, sharper spectral features, at $-20$ and $90$\,\kms,
each comprising several velocity components. Despite the complexity,
no substantial differences were required between the velocity
structures fitted to the Zn and \Ion{Cr}{ii} transitions. As with
J1029$+$1039, the significant blending from \tran{Mg}{i}{2026} is
strongly constrained with the high \SN\ spectrum of
\tran{Mg}{i}{2852}. There is no evidence for problems with the
individual residual spectra or the CRS in \Fref{f:J0226}, and the
final $\chi^2_\nu$ in \Tref{t:res} satisfies our criterion in
\Sref{ss:genfit}. We therefore accepted this absorber into our
main results.

\subsubsection{$\zab=1.971$ towards Q1755$+$57}\label{sss:Q1755}

\begin{figure*}
\begin{center}
\includegraphics[width=0.85\textwidth]{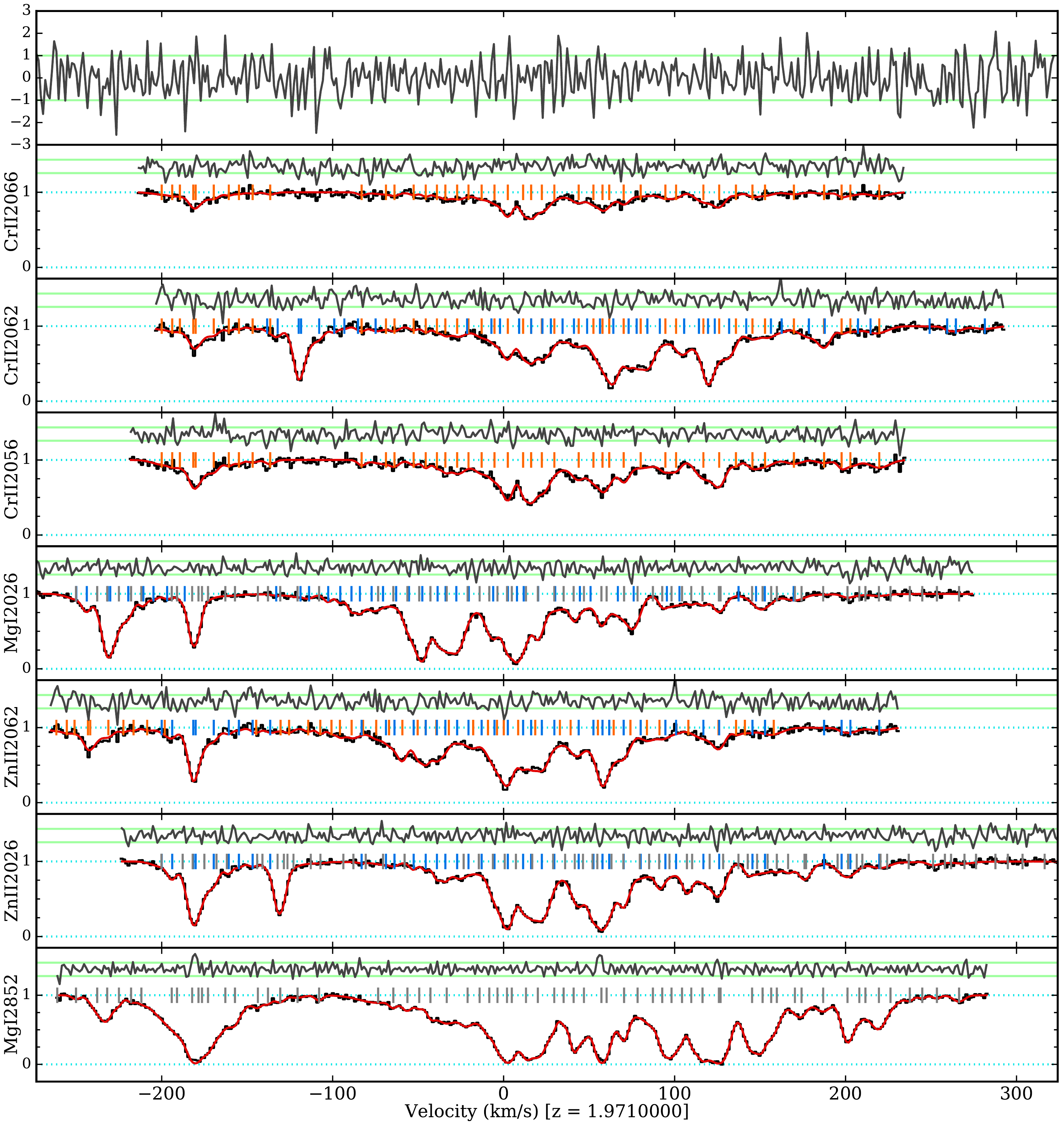}\vspace{-1em}
\caption{Same as \Fref{f:J0058H} but for the Keck/HIRES spectrum of the $\zab=1.971$ absorber towards Q1755$+$57.}
\label{f:Q1755}
\end{center}
\end{figure*}

This absorber is significantly more complex than any other in this
study. The Keck/HIRES spectrum in \Fref{f:Q1755} shows that the
Zn/\Ion{Cr}{ii} absorption spans 450\,\kms\ (and \tran{Mg}{i}{2852}
spans 540\,\kms), with 4 prominent spectral features at $-180$, 10, 70
and 120\,\kms, all of which comprise many velocity components. The
sharpest features in both Zn and \Ion{Cr}{ii} are those at 70 and
120\,\kms\ and these provide the strongest constraints on \daa. It is
interesting that the features at $-180$ and 10\,\kms\ are strong and
sharp in \Ion{Zn}{ii} but much less so in \Ion{Cr}{ii}; this is most
likely due to enhanced levels of dust-depletion in these parts of the
absorber \citep[e.g.][]{Pettini:1990:48,Zych:2009:1429}. The velocity
structure comprises a total of 44 fitted velocity components in
\Ion{Cr}{ii}, with all but 3 of those required to fit the \Ion{Zn}{ii}
absorption. These 3 components sit in the extreme blue and red parts
of the profile, at $-189.5$, 135.8 and 169.7\,\kms, and are all very
weak in \Ion{Cr}{ii}, so their presence in, or absence from the fit to
\Ion{Zn}{ii} has no significant effect on \daa.

As with the other moderately complex absorbers in this study, towards
J1029$+$1039 and J1237$+$0106, fitting the \tran{Mg}{i}{2026} that
blends with \tran{Zn}{ii}{2026} is very important but is also tightly
constrained by the high \SN\ of the \tran{Mg}{i}{2852}
transition. And, even though this absorber is very complex, and the
Voigt profile fitting analysis correspondingly difficult and
time-consuming, the final fit in \Fref{f:J0226} shows no evidence of
significant runs of deviant residuals in individual transitions or the
CRS. The final $\chi^2_\nu$ in \Tref{t:res} is well below our
threshold of 1.2, so this absorber is accepted into our main results.

As mentioned in \Sref{ss:obs}, the spectrum of Q1755$+$57 is the only
one in which the Zn and \Ion{Cr}{ii} transitions are not all covered
by the same exposures. Of the 13 exposures contributing to the HIRES
spectrum, 4 used the 318--605-nm wavelength setting which covers the
\tran{Zn}{ii}{2026} transition but not the redder \Ion{Zn}{ii}
transition or the \Ion{Cr}{ii} triplet. It is therefore possible that
residual slit positioning effects may introduce a small, spurious
shift between \tran{Zn}{ii}{2026} and the other transitions. However,
by constructing a sub-spectrum of Q1755$+$57 using only the 9
exposures in \Tref{t:obs} that cover all the Zn and \Ion{Cr}{ii}
transitions, we find a \daa\ value that differs by $<$0.4\,ppm from
the fiducial one in this absorber, which is $<$10\,per cent of the
statistical uncertainty. The uncertainty is just 5\,per cent
larger. These results indicate that any shift introduced between
\tran{Zn}{ii}{2026} and the other transitions by the 318--605-nm
exposures causes a negligible systematic error in \daa. We therefore
do not add this into the systematic error budget for this absorber in
\Tref{t:res}.

\subsubsection{$\zab=1.921$ towards Q2206$-$1958}\label{sss:Q2206}

\begin{figure}
\begin{center}
\includegraphics[width=0.99\columnwidth]{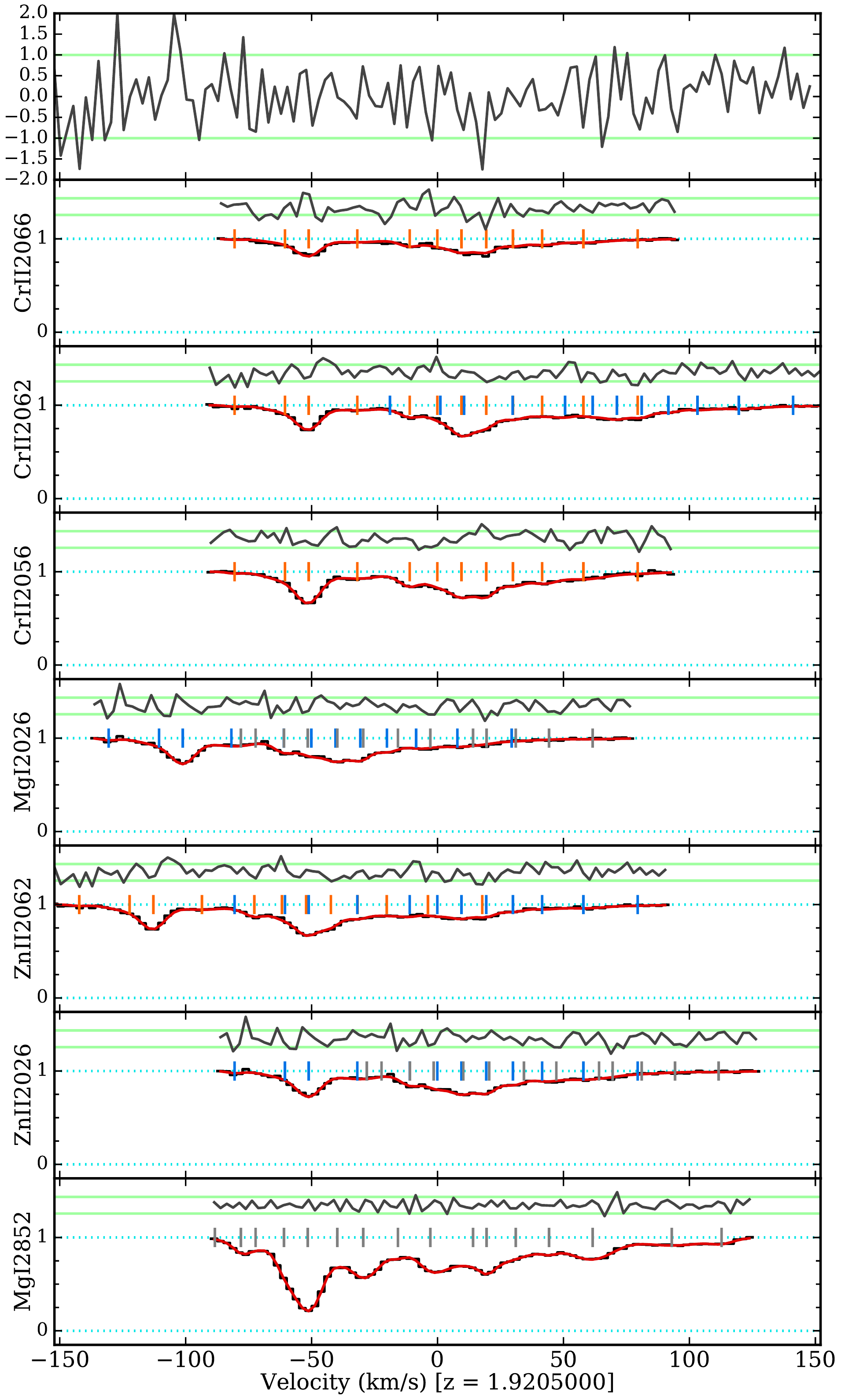}\vspace{-1em}
\caption{Same as \Fref{f:J0058H} but for the VLT/UVES spectrum of the $\zab=1.921$ absorber towards Q2206$-$1958.}
\label{f:Q2206}
\end{center}
\end{figure}

This absorption system was one of the first to be studied in detailed
with Keck/HIRES \citep[e.g.][]{Prochaska:1997:140} and is well-known
to have prominent Zn and \Ion{Cr}{ii} lines. Nevertheless, they are
the weakest in our sample; this absorber was included in this study
because the background quasar is comparatively bright
($V\sim17.3$\,mag) and a large number of archival VLT/UVES exposures
are available (total exposure time, $\sim$15\,h), providing a very
high \SN\ of 97\,pix$^{-1}$ (see \Tref{t:obs}) and, therefore, a small
statistical uncertainty on \daa.

The VLT/UVES spectrum in \Fref{f:Q2206} shows that the Zn and
\Ion{Cr}{ii} absorption spans 160\,\kms\ with two main spectral
features at $-50$ and 10\,\kms. The former is dominated by a single
velocity component and provides the strongest constraint on \daa,
while the latter is more complex, comprising several
components. Therefore, the blend from \tran{Mg}{i}{2026} in
\tran{Zn}{ii}{2026} does not strongly affect the part of the absorber
most important for measuring \daa. In any case, the \Ion{Mg}{i}
velocity structure is strongly constrained by the \tran{Mg}{i}{2852}
absorption. The residuals for individual transitions and the CRS in
\Fref{f:Q2206} show no evidence of data artefacts or unfitted
structure, and the final $\chi^2_\nu$ value in \Tref{t:res} is 1.06,
thereby satisfying our selection criteria. This absorber is therefore
accepted into our main results.

\section{Results}\label{s:res}

\subsection{\boldmath{$\daa$} measurements and statistical uncertainties}\label{ss:dares}

\Tref{t:res} shows the best-fit values of \daa, their 1$\sigma$
statistical uncertainties and the final $\chi^2_\nu$ resulting from
the $\chi^2$-minimization analyses of the 12 profile fits in Figs.\ \ref{f:J0058H}--\ref{f:Q2206}. As
discussed in \Sref{ss:fits}, all but one (from the VLT/UVES spectrum
of J0058$+$0041) of the 12 fits pass our selection criteria for
ensuring reliable \daa\ measurements. We therefore only consider the
other 11 \daa\ measurements in the remaining analysis. However, none
of our conclusions would change if the one rejected measurement was
included. The $\left|\daa\right|$ values in \Tref{t:res} for these 11
measurements are all $<$8.5\,ppm, with an average statistical uncertainty
of 6.9\,ppm. The uncertainties range between 2.7 and 12.9\,ppm. One
absorbers has an uncertainty $<$3\,ppm, which is comparable with
recent, high-precision measurements of individual systems which
utilize all transitions, not just Zn and \Ion{Cr}{ii}
\citep[e.g.][]{Molaro:2013:A68,Evans:2014:128}.

Using only the statistical errors in \Tref{t:res}, the weighted mean
\daa\ is $\left<\daa\right>_{\rm w}=1.12\pm1.67$\,ppm, which is
clearly consistent with no cosmological variation in $\alpha$. Note
that this precision is similar to that of the large absorber samples
from Keck and VLT
\citep[$\approx$1.2\,ppm;][]{Webb:2001:091301,Murphy:2004:131,Webb:2011:191101,King:2012:3370}
and somewhat better than that of the 23-absorber VLT sample of
\citet{Chand:2004:853} reanalysed by \citet{Wilczynska:2015:3082},
$\approx$2.3\,ppm. However, the uncertainties quoted for these
previous studies were not just statistical and included some
systematic effects. The $\chi^2$ of the \daa\ values around our
weighted mean -- again, considering only statistical errors -- is 0.69
per degree of freedom, with an associated probability (that $\chi^2$
should be larger by random chance alone) of 74\,per cent. Assuming
that \daa\ does not vary significantly between absorbers, this
indicates that additional, systematic errors that have random
magnitude and sign in individual absorbers, are unlikely to be as
large or larger than the statistical errors. If they were, we should
observe a larger scatter in \daa\ than expected from the statistical
errors alone, and a larger $\chi^2$ value. Nevertheless, we consider
and quantify the most important sources of systematic errors in the
next Section below before further analysing the \daa\ sample in
\Sref{ss:mainres}.

\subsection{Systematic uncertainties}\label{ss:syserr}

\subsubsection{Long-range distortions}\label{sss:long}

The main advantage of using only the Zn and \Ion{Cr}{ii} transitions
to measure \daa\ is that they are very sensitive to $\alpha$ variation
while, by virtue of their wavelength proximity, insensitive to
long-range distortions of the wavelength scale. Indeed, as discussed
in \Sref{s:intro}, a distortion of 0.2\,\ms\,\AA$^{-1}$, which is
typical of those found in recent studies, would only cause a spurious
shift of 16\,\ms\ between the \tran{Zn}{ii}{2026} and
\tran{Cr}{ii}{2062} transitions, and \Eref{e:daa} implies that this
would cause a systematic error in \daa\ of only 0.3\,ppm per
absorber. However, this simple estimate will be conservative because
\tran{Zn}{ii}{2026} is bluewards of all three \Ion{Cr}{ii}
transitions, while \tran{Zn}{ii}{2062} is redwards of two and
bluewards of the other. This more complicated pattern will decrease
the systematic error in \daa\ for a given, assumed distortion.

For the 4 absorbers in \Tref{t:res} with the smallest statistical
errors in \daa, we tested the effect of introducing a
0.2\,\ms\,\AA$^{-1}$ distortion into each exposure, recombining these
distorted exposures in {\sc uves\_popler}, and minimizing $\chi^2$
again using the final profile models. The effect is 0.15--0.25\,ppm,
smaller than the 0.3\,ppm expected from the simple estimate above. In
previous studies where all transitions were used in the MM analysis,
long-range distortions of this magnitude were found to cause
systematic errors in \daa\ of $\sim$5--10\,ppm
\citep[e.g.][]{Evans:2014:128}. This demonstrates the resistance to
long-range distortions of the \Ion{Zn/Cr}{ii} approach.

Below we find that other systematic errors in our analysis are almost
an order of magnitude larger than those estimated from long-range
distortions above. We therefore ascribe the simple, possibly
conservative estimate of 0.3\,ppm to the systematic error budget of
each absorber from this effect. This is reflected in
\Tref{t:res}. When deriving a weighted mean \daa\ for all absorbers
(\Sref{ss:mainres} below), we further assume that the magnitude and
sign of the long-range distortions may be treated as random from
spectrum to spectrum. Figures 5 and 8 of \citet{Whitmore:2015:446}
indeed show significant variation, but the variation is complex and
not understood, so we expect this assumption to be only partially
true. Nevertheless, that this systematic error is relatively small in
our \Ion{Zn/Cr}{ii} analysis means that neglecting this complication
will have a negligible effect on the overall systematic error budget.

\subsubsection{Intra-order distortions}\label{sss:intra}

While the proximity of the Zn and \Ion{Cr}{ii} transitions to each
other makes their exclusive use insensitive to long-range distortions,
it does leave this approach vulnerable to short-range, intra-order
distortions. The \Ion{Cr}{ii} triplet and \tran{Zn}{ii}{2062}
typically span only half the free spectral range of a HIRES or UVES
echelle order, while \tran{Zn}{ii}{2026} typically falls one or two
echelle orders bluewards of these. Therefore, for a single absorption
system, the magnitude of the velocity shifts generated between these
transitions, and so the systematic effect on \daa, clearly depends on
the amplitude of the distortion, the shape of the distortion pattern
repeated from order to order, and where the Zn and \Ion{Cr}{ii}
transitions fall within their respective orders. The effect on \daa\
will be random in sign and magnitude for different absorbers, so it
will tend to average to zero for a large sample. However, for a modest
sample of 11 measurements, the effect is important to assess.

Our approach to testing the sensitivity of \daa\ to this effect is to
apply a `saw-tooth' velocity distortion to each echelle order's
wavelength scale, for all exposures of every quasar in our
sample. That is, a velocity shift of $\Delta v=200$\,\ms\ is applied
to the centre of each order, falling away linearly to $-\Delta v$
at the order edges. Once the distorted orders are
recombined with {\sc uves\_popler}, \daa\ is re-measured in each
absorber using its final profile model as a starting point for the
$\chi^2$ minimization procedure. A similar approach was used in
several other recent varying-constant studies
\citep[e.g.][]{Malec:2010:1541,Molaro:2013:A68,Bagdonaite:2014:10,Evans:2014:128},
with the latter three using smaller amplitudes of $\Delta v=100$\,\ms\
which are typical of those found in both HIRES and UVES spectra in
recent studies \citep{Whitmore:2010:89,Whitmore:2015:446}. However,
because this is the largest systematic effect for our study, we use a
larger amplitude to ensure it includes most, if not all potential
distortions in the quasar spectra.

We find that the mean deviation in \daa\ caused by introducing the
`saw-tooth' intra-order distortions is 2.03\,ppm and, as expected, the
magnitude and sign varies randomly from absorber to absorber. Given
that we do not know the true shape of the intra-order distortion
pattern or its amplitude for any individual absorber, we ascribe this
mean value to the systematic error budget for each absorber. This is
reflected in \Tref{t:res}.

\subsubsection{Spectral redispersion}\label{sss:redisp}

The exposures of each quasar were redispersed onto a common log-linear
wavelength grid, a process that slightly distorts the absorption line
shapes and introduces small correlations in the flux and flux
uncertainties of neighbouring pixels. That is, the redispersion
process will cause small, spurious velocity shifts between the
absorption profiles of different transitions, and hence a systematic
error in \daa\ that varies in sign and magnitude from absorber to
absorber.

As in other recent works
\citep[e.g.][]{Malec:2010:1541,Weerdenburg:2011:180802,Molaro:2013:A68,Evans:2014:128},
we quantify this effect by slightly altering the dispersion of the
final wavelength grid, by $\pm$0.01 and $\pm$0.02\,\kms, for each
quasar and re-measuring \daa\ of each absorber using its final profile
model for the $\chi^2$ minimization. The mean deviation of these 4
measurements from the fiducial \daa\ value indicates the systematic
uncertainty for each absorber, $i$, $\sigma_{\rm disp}^i$. However,
with only 4 measurements, this approach will underestimate the effect
for some absorbers (and overestimate it for others). To address this,
we first noted an expected correlation between $\sigma_{\rm disp}$ and
the statistical uncertainty on \daa, $\sigma_{\rm stat}$. A linear
least-squares fit yields the relationship
$\sigma_{\rm disp}[{\rm ppm}]\approx 0.15+0.12\times\sigma_{\rm
  stat}[{\rm ppm}]$.
Therefore, for absorber $i$, we assigned the larger of the original
estimate, $\sigma_{\rm disp}^i$, and the value derived from this
relationship, as the systematic error from redispersion for that
absorber.

\Tref{t:res} shows the total systematic error budget for each
absorber. As discussed in Sections \ref{sss:long} \& \ref{sss:intra}
above, the systematic uncertainty from long-range and intra-order
distortions are taken to be the same for all absorbers, 0.3 and
2.03\,ppm respectively. Only the systematic uncertainty from
redispersion effects varies from absorber to absorber, so for simplicity
we only present the quadrature sum of all systematic errors in
\Tref{t:res}.

\subsubsection{Isotopic abundance variations}\label{sss:iso}

The profile fits to the \Ion{Zn}{ii}, \Ion{Cr}{ii}, and \Ion{Mg}{i}
absorption lines included the measured or calculated isotopic and
hyperfine structures reviewed recently in \citet{Murphy:2014:388} and
the terrestrial relative isotopic abundances
\citep{Rosman:1998:1275}. However, if the relative isotopic abundances
of Zn and/or Cr are different in the absorption clouds, this will
shift the velocity centroids of the absorption features, causing
systematic errors in \daa\ measured from our fits which use the
terrestrial abundances. This is a well-recognized problem
\citep[e.g][]{Murphy:2001:1208,Murphy:2001:1223}, especially for fits
involving \Ion{Mg}{i/ii} transitions because the separations between
isotopic components are relatively large (because Mg is a relatively
light atom). Indeed, \citet{Fenner:2005:468} found that if the heavy
isotopes of Mg ($^{25,26}$Mg; total relative terrestrial abundance of
21\,per cent) were completely absent in quasar absorption clouds, a
systematic error in \daa\ of $\approx$4.5\,ppm would be induced in a
MM analysis of the strong \Ion{Mg}{i/ii} and \Ion{Fe}{ii} transitions.

However, the \Ion{Zn/Cr}{ii} combination is more resistant to this
systematic error, mainly because of the smaller isotopic shifts in
these heavier atoms. The \Ion{Mg}{i} transitions are also decoupled
from \daa\ in our fits (see \Sref{ss:genfit}), so variations in the Mg
isotopic abundances will not affect our results. Using a chemical
evolution model for spiral galaxies like our Milky Way,
\citet{Fenner:2005:468} predicted that the isotopes of Cr and Zn with
the lowest terrestrial abundances will be less abundant in
gas with low metallicities typical of most quasar absorbers. And while
our absorbers will have higher-than-average metallicities -- they were
selected because of their very strong metal absorption -- and this may
reduce the effect on \daa, we cannot rule out this systematic effect
on such grounds.

If we make the extreme assumption that the sub-dominant isotopes are
absent in the absorbers, our fits will find centroids for the
\Ion{Zn}{ii} and \Ion{Cr}{ii} transitions that are too red by
$\approx$100\,\ms\ and too blue by $\approx$10\,\ms,
respectively. Considering just the \tran{Zn}{ii}{2026} and
\tran{Cr}{ii}{2056} transitions, this would cause a systematic error
in \daa\ of $-2.3$\,ppm (cf.\ 4.5\,ppm for the \Ion{Mg}{i/ii} and
\Ion{Fe}{ii} combination). Just as for the long-range distortions
(\Sref{sss:long}), this effect will be diminished by
\tran{Zn}{ii}{2062} falling amongst the \Ion{Cr}{ii} triplet. Indeed,
in our best-constrained absorber, that in the Keck spectrum of J0841$+$0312, we find
that \daa\ shifts by $-1.9$\,ppm if we remove the sub-dominant
isotopes from the fit. This is similar to our total
error budget of 1.9\,ppm for the ensemble of 11 measurements; correcting
for even this extreme possibility would place our measured \daa\ value
just 1.6$\sigma$ above zero.

Given that \daa\ is relatively insensitive to this effect, our general
lack of information about the isotopic abundances in the absorbers,
and following all previous measurements which include the more
sensitive \Ion{Mg}{i/ii} transitions, we do not attempt to formally
include this effect in our systematic error budget. However, our main
conclusions in \Sref{s:conc} are explicitly stated with the assumption
of terrestrial isotopic abundances.

\subsection{Statistical analysis of main results}\label{ss:mainres}

\begin{figure}
\begin{center}
\includegraphics[width=0.99\columnwidth]{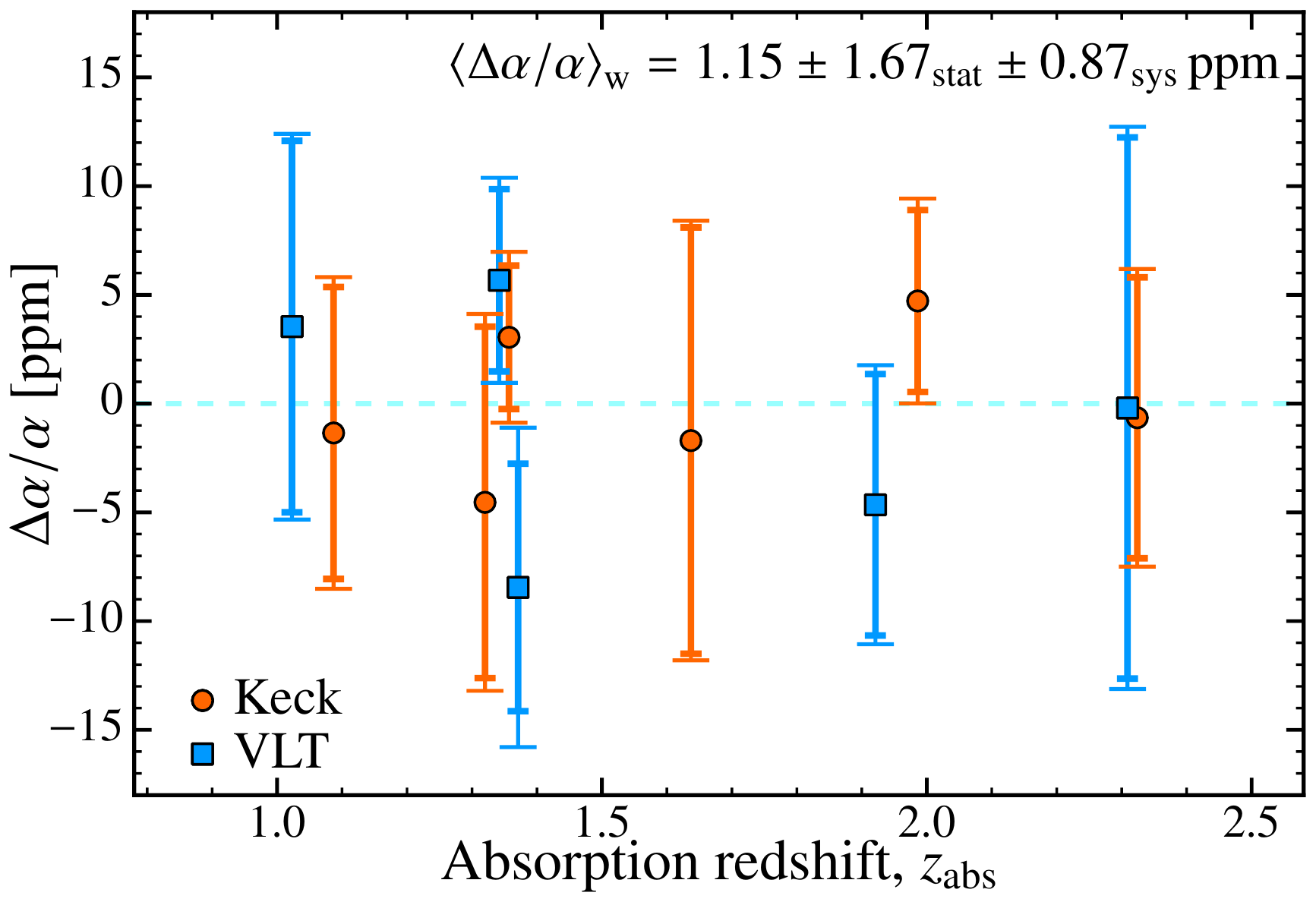}\vspace{-1em}
\caption{The \daa\ measurements for the 11 profile fits that pass our
  selection criteria (i.e.\ all but the VLT spectra of J0058+0041 from
  \Tref{t:res}) versus their absorption redshifts, \zab. The thicker,
  shorter error bars represent the 1$\sigma$ statistical uncertainties
  and the thinner, longer ones represent the quadrature sum of
  the statistical and systematic uncertainties (see \Tref{t:res}). The
  values derived from Keck spectra are shown as circles, and values
  from VLT as squares. Note that the Keck results are all shifted by
  $\Delta z=0.015$ for clarity. The weighted mean \daa\ from the 11
  measurements, $\left<\daa\right>_{\rm w}$ [\Eref{e:wmean}], is also
  shown. These results were derived using the terrestrial isotopic abundances for Zn and Cr; see discussion of this potential systematic error in \Sref{sss:iso}.}
\label{f:da_vs_z}
\end{center}
\end{figure}

\Tref{t:res} shows the best-fit value of \daa, the 1$\sigma$
statistical uncertainty, and systematic uncertainty for all absorbers
in our study. Note that, as described in \Sref{sss:J0058}, the result
from the VLT spectrum of J0058+0041 (italics in \Tref{t:res}) is not
considered further in our analysis because the fit in that case
returned a $\chi^2_\nu$ value exceeding our threshold value of
1.2. The remaining 11 \daa\ values and uncertainties are plotted in
\Fref{f:da_vs_z} against absorption redshift and delineated by
telescope. It is immediately clear that all 11 values are consistent
with no variation in $\alpha$ and that there is no systematic offset from $\daa=0$. There is no significant trend with
redshift, in either the sample as a whole or from either telescope
separately. All 11 results are also consistent with each other. In
particular, a reliable \daa\ measurement was made with both Keck and
VLT spectra of the same two absorbers towards PHL957 and J0841$+$0312,
and these are found to be consistent with each other.

As discussed in Sections \ref{sss:intra} \& \ref{sss:redisp}, the two
largest systematic effects on \daa\ -- those from intra-order
distortions and spectral redispersion -- are random from absorber to
absorber. Therefore, we combined the results to form a weighted mean
in a straight-forward way: the weight for each absorber, $i$, is the
inverse sum of the statistical and systematic variances,
i.e.\
$w_i\equiv(\sigma_{i,{\rm stat}}^2+\sigma_{i,{\rm sys}}^2)^{-1}$; the
statistical uncertainty is
$\sigma_{\rm stat}=1/\sqrt{\sum_i{1/(\sigma_{i,{\rm stat}})^2}}$; and
the systematic uncertainty is the quadrature difference between
$\sigma_{\rm tot}=1/\sqrt{\sum_iw_i}$ and $\sigma_{\rm stat}$. This
provides our main result: the weighted mean \daa\ from the 11
measurements, with 1$\sigma$ statistical and systematic uncertainties:
\begin{equation}\label{e:wmean}
\left<\daa\right>_{\rm w} = 1.15 \pm 1.67_{\rm stat} \pm 0.87_{\rm sys}\,{\rm ppm}\,.
\end{equation}
The scatter around this weighted mean is consistent with expectations
from the individual error bars: $\chi^2=5.04$ which, for 10 degrees of
freedom, has an associated probability (of being exceeded by chance
alone) of 89\,per cent. The weighted means for the Keck and VLT
sub-samples are $1.7\pm2.1_{\rm stat}\pm 1.1_{\rm sys}$ (6 absorbers)
and $0.3\pm2.7_{\rm stat}\pm 1.4_{\rm sys}$\,ppm (5 absorbers),
respectively; these are consistent with each other within
0.4$\sigma_{\rm comb}$ where $\sigma_{\rm comb}$ is the quadrature sum
of the statistical and systematic uncertainties. The scatter amongst the VLT
measurements is consistent with the uncertainties ($\chi^2=3.4$ around
the VLT weighted mean, with associated probability of 49\,per
cent). The scatter amongst the 6 Keck results may seem too small,
relative to the uncertainties, in \Fref{f:da_vs_z}: $\chi^2=1.5$
around the Keck weighted mean. However, for such a small sample, this
is expected to occur 8\,per cent of the time by chance alone, so
there is no strong evidence for an underestimated scatter (or
overestimated uncertainties).

\section{Discussion}\label{s:disc}

Our new measurement of \daa\ in \Eref{e:wmean} is the first from a
statistical sample of absorbers that is resistant to long-range distortions at the $<$1\,ppm level. \daa\
has been measured in just four other distinct absorbers with
distortion-corrected spectra, by \citet{Evans:2014:128} and
very recently by \citet{Kotus:2016:arXiv}, using MM analyses of a large variety of
transitions. \citet{Evans:2014:128} studied 3 absorbers towards a
single quasar (HS1549$+$1919; $\zab=1.143$, 1.342 and 1.802) using 3
different telescopes (Keck, Subaru and VLT). The weighted mean from
their 9 measurements is
$\left<\daa\right>_{\rm w} = -5.4 \pm 3.3_{\rm stat} \pm 1.5_{\rm sys}$\,ppm.
\citet{Kotus:2016:arXiv} analysed the very complex
$\zab=1.1508$ absorber towards HE0515$-$4414 (the brightest known
southern quasar at $\zem>1$) using very high-quality VLT/UVES spectra
($\SN\sim250$\,per 1.3-\kms\ pixel). They obtained the most precise
measurement in a single absorber to date:
$\daa = -1.4 \pm 0.6_{\rm stat} \pm 0.6_{\rm sys}$\,ppm. These 2 distortion-corrected
results are consistent with each other and with our new distortion-resistant measurement, and the combined constraint on
\daa\ is a weighted mean [derived with the same approach as
\Eref{e:wmean}] of
\begin{equation}\label{e:all}
\left<\daa\right>_{\rm w} = -1.2 \pm 0.5_{\rm stat} \pm 0.6_{\rm sys}\,{\rm ppm}\,.
\end{equation}
This is the best current limit on variations in $\alpha$ using quasar absorption systems, assuming that it does not vary across the sky.

Given that the quasars in our sample have a reasonable distribution
across the sky, testing whether $\alpha$ displays such spatial
variation is possible, in principle. However, the sample is still
relatively small (9 quasars) so, instead, a comparison with recent,
possible evidence for such variations is more instructive. Prior to
long-range distortions being found in UVES and HIRES spectra, the
large statistical Keck and VLT samples were combined to search for
angular variations in \daa\
\citep{Webb:2011:191101,King:2012:3370}. The simplest model of such
variations, a dipole, was found to be required at $\approx$4$\sigma$
significance, with the pole in the direction
${\rm (RA, Dec.)}=(17.4\pm0.9{\rm \,h}, -58\pm9^\circ)$ and deviating
by $\daa = 10.2^{+2.2}_{-1.9}\cos(\Theta)$\,ppm with the angular
separation, $\Theta$, from this direction
\citep{King:2012:3370}. \citet{Whitmore:2015:446} found that
long-range distortions likely explain the VLT results and, at least
partially, the Keck results that provide a basis for this dipole
result. Nevertheless, new measurements that are resistant to, or
corrected for, the long-range distortions can in principle be used to
rule out or confirm this variation in $\alpha$ across the sky.

\begin{figure*}
\begin{center}
\centerline{\hbox{
    \includegraphics[width=1.05\columnwidth]{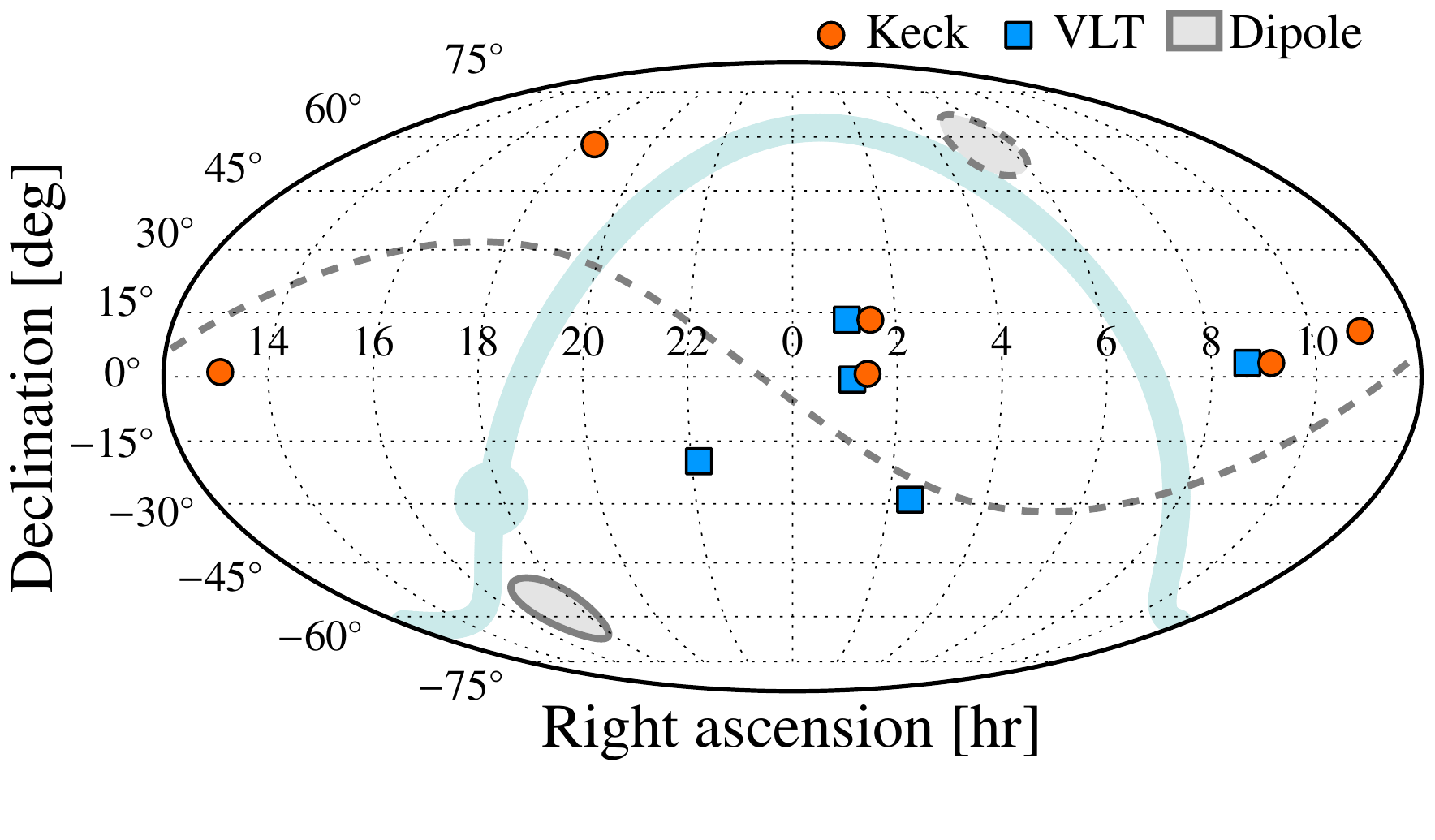}
    \hspace{0.01\columnwidth}
    \includegraphics[width=0.94\columnwidth]{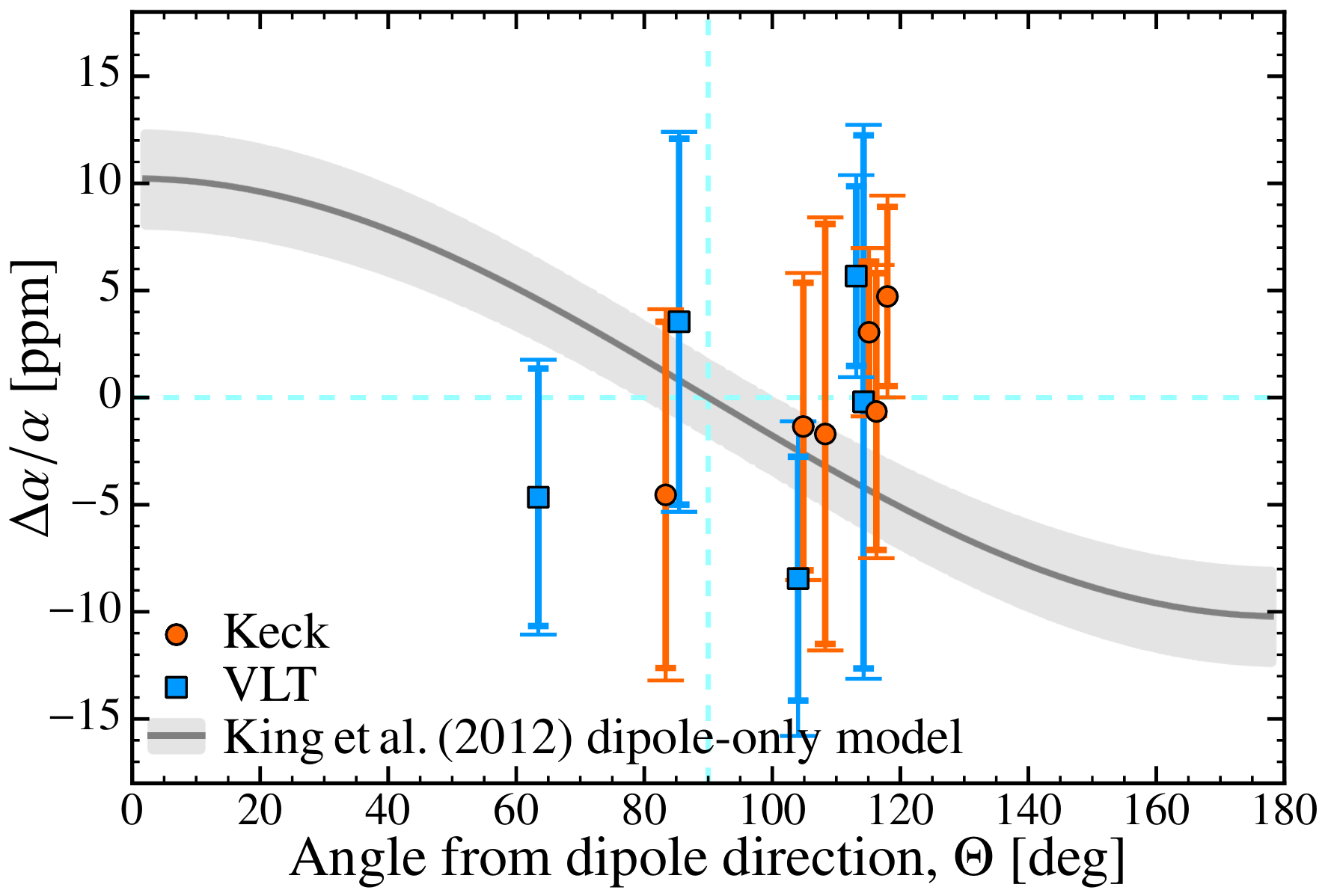}
}}
\caption{Left panel: Angular distribution of our 9 quasar lines of
  sight across the sky. The quasars observed with Keck are shifted by
  $\Delta{\rm RA}=0.5$\,h for clarity. The shaded ellipse with the solid line represents
  the 1$\sigma$ uncertainty in the position of the pole in the
  dipole-only model of \daa\ from \citet{King:2012:3370}; the ellipse with the dashed line represents the anti-pole and the dashed curve is the dipole's equator. The lighter
  blue shading represents the Galactic plane, with the bulge
  indicating the Galactic centre. Right panel: Comparison of our 11 new
  \Ion{Zn/Cr}{ii} measurements of \daa\ with the dipole expectation
  (solid line; shading indicates the $\pm$1$\sigma$ uncertainty). Our
  quasar sight-lines are distributed near the equator of the dipole
  model, where \daa\ is predicted to be $<$5\,ppm, so we cannot rule
  out or confirm the dipole with our new measurements alone.  }
\label{f:da_vs_theta}
\end{center}
\end{figure*}

The left panel of \Fref{f:da_vs_theta} compares the distribution of
our quasar sight-lines on the sky with the position of the putative
dipole, and the right panel compares our new \daa\ measurements with
the predicted value from the dipole model. It is immediately clear
that our quasar sight-lines are positioned around the equator of the
dipole model, with angular separations clustered near
$\Theta\approx90^\circ$ where the model predicts $\daa<5$\,ppm. This
precludes an effective test of the dipole model using our new
\Ion{Zn/Cr}{ii} measurements. Indeed, the $\chi^2$ between our values
and the model is 14.9 which, for 11 degrees of freedom, has an
associated probability of 18\,per cent, i.e.\ the data are consistent
with the model. The distortion-corrected measurements from
HE0515$-$4414 \citep{Kotus:2016:arXiv} and HS1549$+$1919
\citep{Evans:2014:128} have $\Theta$ values of 78 and $80^\circ$,
respectively, so adding them to this comparison would not provide
significant additional discriminating power. The simplest approach to
rule out or confirm the dipole model would be to obtain a modest
sample of distortion-correct spectra and/or strong \Ion{Zn/Cr}{ii}
absorbers, towards quasars within $\sim$20$^\circ$ of the pole
and/or anti-pole.

Our new measurements demonstrate the important advantages of using
only the Zn and \Ion{Cr}{ii} transitions to measure \daa, primarily
their high sensitivity to $\alpha$-variation and resistance to
long-range distortions in the wavelength calibration of the quasar
spectra. However, there are some disadvantages of this
approach. Firstly, there are only 2 strong \Ion{Zn}{ii} transitions
and 3 strong \Ion{Cr}{ii} transitions, so each \daa\ measurement
relies on fewer transitions than is typical for most MM analyses,
which reduces the statistical precision available. Secondly, they are
normally weak in most absorption systems, so large samples of
high-resolution quasar spectra are needed before a statistical sample
of strong-enough \Ion{Zn/Cr}{ii} absorbers can be identified towards
bright-enough quasars. Pre-selection of targets could be made
efficiently from short, moderate-resolution ($R\sim10000$) follow-up
spectra of absorbers which show possible Zn and \Ion{Cr}{ii}
absorption in low-resolution spectra from large surveys (e.g.\ from the Sloan Digital Sky
Survey). Thirdly, the blends between the \tran{Mg}{i}{2026} and
\tran{Zn}{ii}{2026} transitions, and to some extent the fact that
\tran{Cr}{ii}{2062} and \tran{Zn}{ii}{2062} overlap in absorbers with
extended velocity structure, complicates the analysis. We have
demonstrated here how that can be overcome by using \tran{Mg}{i}{2852}
to constrain the \tran{Mg}{i}{2026} blend, with the two transitions `decoupled' in the fitting procedure
(see \Sref{ss:genfit}). Despite these disadvantages, we have demonstrated that \Ion{Zn/Cr}{ii} measurements of \daa\ offer an
important complement to other MM analyses: the latter may, in principle, achieve higher
statistical precision but they are more vulnerable to
long-range instrumental systematic effects than the \Ion{Zn/Cr}{ii} approach.

\section{Conclusions}\label{s:conc}

Comparing the two \Ion{Zn}{ii} and the three \Ion{Cr}{ii} transitions
falling in the narrow rest wavelength range of 2026--2066\,\AA\ is, in
principle, the most sensitive to $\alpha$-variation of all metal line
combinations used in many-multiplet studies to date
(\Fref{f:q_vs_wl}). However, these transitions are rarely strong
enough, even in damped Lyman-$\alpha$ systems, to be very useful in
constraining \daa. We identified 9 absorption systems at
$\zab=1.0$--2.4 with strong \Ion{Zn/Cr}{ii} absorption towards 9
relatively bright quasars ($r=16.1$--18.3\,mag) and obtained new
and/or archival spectra with Keck/HIRES (6 quasars) and VLT/UVES (6
quasars, 3 in common with Keck) with high \SN\ (mean
54\,pix$^{-1}$ around \Ion{Zn/Cr}{ii}). The quasar spectra and \Ion{Zn/Cr}{ii} absorption
profile fits are publicly available in
\citet{Murphy:2016:alphaZnCrII2016}. Of the 12 profile fits (Figs.\
\ref{f:J0058H}--\ref{f:Q2206}), 11 passed our stringent selection
criteria to provide robust new measurements of \daa. These all lie
within 8.5\,ppm of zero, with total (quadrature addition of
statistical and systematic) uncertainties of 3.5--13.2\,ppm, and all
formally consistent with $\daa=0$. There is no evidence of redshift
evolution in \daa\ or any systematic difference between the Keck and
VLT measurements. The final weighted mean \daa\ from all 11
measurements, with 1$\sigma$ statistical and systematic uncertainties,
is $1.2\pm1.7_{\rm stat}\pm0.9_{\rm sys}$\,ppm [1$\sigma$, \Eref{e:wmean}],
consistent with no variation in $\alpha$. As with other many-multiplet measurements of \daa, this assumes that the relative isotopic abundances of Zn and Cr in the absorbers are similar to the terrestrial values. Assuming that isotopes with the lowest terrestrial abundances are completely absent in the absorbers, \daa\ would be corrected to $3.1\pm1.9$\,ppm, still consistent with no variation in $\alpha$.

An important advantage of the \Ion{Zn/Cr}{ii} approach demonstrated
here is its resistance to long-range distortions in the wavelength
calibration of the quasar spectra. Such systematic effects will have
significantly impacted all previous \daa\ measurements with the
many-multiplet method, except for the recent analyses of
distortion-correct spectra by \citet{Evans:2014:128} and
\citet{Kotus:2016:arXiv}. Indeed, these distortions likely explain
previous evidence for $\alpha$-variation from large Keck and VLT
absorber samples \citep{Whitmore:2015:446}. However, the wavelength
proximity of the Zn and \Ion{Cr}{ii} transitions, and their high
sensitivity to $\alpha$-variation, mean that long-range distortions
only cause a $\sim$0.3\,ppm systematic error in each absorber. The
largest systematic error in our \Ion{Zn/Cr}{ii} measurements --
$\sim$2.0\,ppm for each absorber -- stems, instead, from intra-order
distortions. Errors from the redispersion of the quasar spectra
(necessary when combining many exposures of a single quasar together)
cause errors of 0.5--2.3\,ppm in each absorber, evidently scaling with
its statistical uncertainty.

Our new measurements are consistent with the only distortion-corrected
measurements so far, with a total ensemble uncertainty (1.9\,ppm;
quadrature sum of final statistical and systematic uncertainties)
approximately half that from the 9 measurements by
\citet[][3.6\,ppm]{Evans:2014:128} and twice that obtained from
the single, very high-\SN\ absorber of
\citet[][0.9\,ppm]{Kotus:2016:arXiv}. Combined, these measurements
provide the best current restriction on variations in $\alpha$, with a
weighted mean $\daa=-1.2\pm0.5_{\rm stat}\pm0.6_{\rm sys}$ [1$\sigma$,
\Eref{e:all}] at redshifts $\zab=1.0$--2.4. Despite this precision,
the spatial distribution of the quasars precludes a rigorous test of
the dipole-like variation in $\alpha$ across the sky from the large
Keck and VLT samples in \citet{King:2012:3370} (i.e.\ taking evidence
for that at face value, ignoring the concerns about long-range
distortions raised above). New \daa\ measurements at a wider variety
of sky coordinates, using distortion-corrected spectra and/or the
\Ion{Zn/Cr}{ii} approach, would enable such a test.

\section*{Acknowledgements}

We thank Sebastian Lopez for initial discussions about using Zn and
\Ion{Cr}{ii} to measure \daa, and Julija Bagdonaite and Jonathan Whitmore for assistance with \Fref{f:da_vs_theta}. MTM thanks the Australian Research
Council for \textsl{Discovery Projects} grant DP110100866 which
supported this work. Some of the data presented herein were obtained
at the W.M.\ Keck Observatory, which is operated as a scientific
partnership among the California Institute of Technology, the
University of California and the National Aeronautics and Space
Administration. The Observatory was made possible by the generous
financial support of the W.M.\ Keck Foundation. The authors wish to
recognize and acknowledge the very significant cultural role and
reverence that the summit of Maunakea has always had within the
indigenous Hawaiian community.  We are most fortunate to have the
opportunity to conduct observations from this mountain. Other data
presented herein were based on observations made with ESO Telescopes
at the La Silla Paranal Observatory under programme IDs 65.O-0158,
67.A-0022, 072.A-0346, 074.A-0201, 079.A-0600, 082.A-0682, 082.A-0569,
083.A-0874 and 084.A-0136.















\section*{Erratum}\label{s:serr}

In MNRAS, 461, 2461 (2016, arXiv:1606.06293v1), the values presented for the relative deviation in the fine-structure constant from the current laboratory value, \daa, resulted from profile fits used to produce the plots of the absorption lines in each system (figures 2--13), not the fiducial fits as intended. For simplicity and clarity in those plots, the fits omitted the isotopic structures of the \Ion{Zn}{ii}, \Ion{Cr}{ii} and \Ion{Mg}{i} transitions, and omitted the very weak \tran{Cr}{ii}{2026} transition. These have generally negligible effects on the values of \daa\ ($\la$0.3\,ppm). However, a separate oversight meant the result for the $\zab=1.371$ absorber towards quasar J0108$-$0037 was mistakenly taken from a fit used for checking for additional velocity components. This resulted in a substantially smaller statistical uncertainty on \daa\ than from the fiducial fit (2.74\,ppm instead of 5.69\,ppm). Because this absorber is amongst the strongest in our sample, and because it has a relatively narrow and simple velocity structure, its value of \daa\ is most strongly affected by \tran{Cr}{ii}{2026.27} which weakly blends with the important \tran{Zn}{ii}{2026.14} transition. This blend was not discussed in the original paper, even though it was used in all the fiducial fits, because we found it to have such a small effect on \daa: fits with and without \tran{Cr}{ii}{2026} included gave values differing by a median (mean) of just 0.05 (0.16)\,ppm. However, the same comparison using the correct fiducial fit for J0108$-$0037 now reveals a difference of 2.5\,ppm.

The correct set of results is now used throughout this version of the paper (i.e.\ arXiv:1606.06293v2), including all numerical results, figures and tables. However, the general conclusions of the original paper remain unchanged. An erratum has been accepted to MNRAS.

\bsp	
\label{lastpage}
\end{document}